\documentclass{article}
\usepackage{graphicx} 
\usepackage[english]{babel}
\usepackage[bottom=3.5cm, top=3.5cm]{geometry}
\usepackage{eurosym}
\usepackage{setspace}
\usepackage{appendix}
\usepackage{placeins} 
\usepackage{todonotes}
\usepackage{natbib}
\usepackage{booktabs}
\usepackage{multirow}
\usepackage{xcolor}
\usepackage{bm}

\usepackage{fullpage}
\usepackage{xurl}

\usepackage [autostyle, english = american]{csquotes} 
\MakeOuterQuote{"} 

\graphicspath{ {./figs} }

\title{Human foraging strategies flexibly adapt to resource distribution and time constraints}

\date{}

\author{Valeria Simonelli$^{*a,b}$, Davide Nuzzi$^{*a}$, Gian Luca Lancia$^a$, Giovanni Pezzulo$^{\dag a}$  \\
\\
$^a$ Institute of Cognitive Sciences and Technologies, National Research Council, Rome, Italy  \\
$^b$ University of Rome “La Sapienza”, Rome, Italy}

\footnotetext{These authors contributed equally to this work}
\footnotetext{Corresponding author: giovanni.pezzulo@istc.cnr.it}

\begin{document}
\maketitle

\begin{abstract}

Foraging is a crucial activity, yet the extent to which humans employ flexible versus rigid strategies remains unclear. This study investigates how individuals adapt their foraging strategies in response to resource distribution and foraging time constraints. For this, we designed a video-game-like foraging task that requires participants to navigate a four-areas environment to collect coins from treasure boxes within a limited time. This task engages multiple cognitive abilities, such as navigation, learning, and memorization of treasure box locations. Findings indicate that participants adjust their foraging strategies -- encompassing both stay-or-leave decisions, such as the number of boxes opened in initial areas and behavioral aspects, such as the time to navigate from box to box -- depending on both resource distribution and foraging time. Additionally, they improved their performance over time by reducing uncertainty about resource locations and distributions, demonstrating enhancements in both foraging strategies and navigation skills. Finally, participants' performance was initially distant from the reward-maximizing performance of optimal agents due to the learning process humans undergo. However, it approximated the optimal agent's performance towards the end of the task, without fully reaching it. These results highlight the flexibility of human foraging behavior and underscore the importance of employing optimality models and ecologically rich scenarios to study foraging.

\end{abstract}


\newpage
\section{Introduction}

Foraging is a nearly universal behaviour in the animal kingdom and essential for survival of most species \citep{mobbs2018foraging,hayden2018economic,hills2006animal,calhoun2015foraging,stephens1986foraging}. Foraging problems encompass not just the search for survival-related resources (e.g., foraging for food) but also the search for information in the environment (e.g. in internet search \citep{Pirolli2007}) or in memory (e.g., "mental search" \citep{hills2015exploration,wilke2009fishing,Todd2020ForagingIM}). 

In natural foraging problems, such as when animals navigate the environment to search for food (or other valuable resources), choice offers are presented serially and often found in clumps or patches, which gradually exhaust. This makes foraging scenarios conceptually different from classical economic tasks used to study value-based decisions, in which choice offers are typically presented simultaneously. Given their peculiarities, foraging choices are often conceptualized in terms of stay or leave decisions; namely, deciding whether to accept the available alternative (e.g., continue foraging in the same patch, in so-called patch-leaving problems) or reject it, in the hope that a better one would be found (e.g., move to another patch) \citep{charnov1976optimal,hayden2018economic,hayden2018neuronal,genovesio2014prefrontal,byrne1988machiavellian}. Stay-or-leave problems are extensively studied, both in humans and animals (e.g., mammals \citep{Blanchard2015,Kane2022,Hayden2011,Gancarz2023,shahidi2024population}, birds \citep{Stephens2009,Chandel2021}). Similarly, the trade-offs between exploration and exploitation that arise during foraging and related tasks have been widely studied in humans and other animals  \citep{Cohen2008,Mehlhorn2015,addicott2017primer,daw2006cortical,verschure2014and,maffei2015embodied,parr2022active,pezzulo2018hierarchical,friston2017active}). 

A formal solution to patch-leaving and other foraging problems is the Marginal Value Theorem of optimal foraging theory \citep{charnov1976optimal}, according to which the most profitable moment to leave a patch ("giving-up time") is when the reward in the current patch is less than the average reward rate of the environment \citep{constantino2015learning}. Various studies reported that humans and many other animal species behave very close, despite not as good as, the optimal behavior suggested by optimal foraging theory \citep{pyke1977optimal,zhang2015using,wolfe2013time,webb2024foraging,garrett2020biased}. This suboptimality has many potential causes; among them, patch leaving decisions are influenced by previous performance \citep{Hutchinson2008,Lin2023} as well as individual \citep{vanDooren2021}, social \citep{Bidari2022}, and aging factors \citep{Lloyd2023}. 

A series of human studies tested how environmental constraints -- the most important being resource distribution -- affect individuals' patch-leaving strategies. Using a multiple-target visual search task, Cain and colleagues \citep{Cain2012} demonstrated that foragers adapt to environmental statistics in a way that is consistent with, although not identical to, the optimal formulation: participants search more when expecting a greater amount of targets and modify their expectations in real time by considering the target distributions across trials. In a different experiment, a virtual patch-foraging task was used to assess the influence of environmental richness on foraging decisions \citep{constantino2015learning}. Results have shown that although participant's behavior was in agreement with the optimal policy there was an "overharvesting" tendency, namely, a tendency to harvest a patch longer than it would be needed for reward maximization; see also \citep{Hutchinson2008}. Overharvesting has been explained as a strategy for reducing the cost of behavioral variability \citep{Cash-Padgett2020}, as a result of optimal learning and adaptability \citep{Harhen2023}, as a result of the the inherent asymmetrical cost, in foregone reward, of understaying, especially at longer travel times \citep{kacelnik1984central}, or as a phenomenon occurring as a consequence of temporal cognition \citep{Kendall2022}. Indeed, time definitely matters when making patch-leaving decisions; as the MVT states, longer travel times lead to longer patch residence times, which lowers the forager's opportunity cost of time as well as the environment's overall rate of reward \citep{bustamante2023effort}.


Like resources are limited in the environment, time is also a finite resource. Animals allocate different time foraging, depending on external constraints and they flexibly adjust their foraging behavior to time available \citep{krebs1980optimal,swennen1989time,ydenberg1994time,ydenberg1998simple}. For example, they tend to work harder when time available for foraging is more limited \citep{lewis2004flexible}. Time pressure also implies that decisions have to be made faster. A recent study employed a time bandit task to examine the effects of time pressure on exploration and decision making, found that in limited time conditions, participants earned less reward, were less sensitive to reward values in repeat choice behavior, were less likely to choose options that were associated with higher uncertainty, and explored less in uncertain scenarios. These findings can be explained by considering that under time pressure, participants tend to select simpler, lower-cost policies \citep{Wu2022}. In sum, the above studies contributed to shed light on how two environmental constraints -- namely, resource distribution and time availability -- affect human patch-leaving strategies. However, these two constraints have been manipulated in isolation, leaving open the question of how they might interact and whether participants are sensitive to both factors. 

Finally, most human foraging studies used simple paradigms, in which rewards are paired to buttons that are simultaneously present on the screen. It is unclear how well their results generalize to foraging scenarios that resemble more closely animal patch-living problems, in which rewards are presented serially, visibility is partial and which engage sophisticated cognitive processes, such as those required to learn and remember reward locations, orienting and navigating in space, etc. \citep{mobbs2018foraging,hayden2018economic,yoo2021continuous,cisek2014challenges}. Crucially, these more challenging and ecologically valid scenarios present various sources of uncertainty about resource locations and distributions—for example, how many reward locations can be reached within limited time, where rewards are spatially located, and how to navigate efficiently to reach them—hence creating multiple occasions for learning. Therefore, they allow for novel questions about how foraging strategies adapt to environmental demands over the experiment, a factor often sidestepped in accounts like optimal foraging theory. For example, they enable researchers to study whether participants improve their foraging strategy and performance over time (both across and within trials) as they reduce uncertainty about key task dimensions, and whether, at some point, they match (or closely approximate) optimal solutions from foraging theory \citep{allen2024using,gordon2021road,maselli2023beyond}.

In this study, we address these challenges by employing a novel patch-leaving task in the form of a video game (Figure \ref{fig:screenshots}). Our research aims to answer three main research questions. First, we aim to examine whether participants flexibly modulate their patch-leaving strategies to different conditions of resource distribution and time availability. Second, we aim to establish whether participants improve their performance as they reduce their uncertainty about the task and whether their putative improvements depend on improved skill (e.g., navigation skill) or also strategic changes. Third, we aim to investigate the degree to which participants patch-leaving behavior under the different conditions align with an optimal foraging agent that maximizes reward.


\section{Methods}

\subsection{Participants and design}

A total of 34 subjects participated to our experiment (age 19-50; 21 female; 13 male; M= 25.3; SD= 5.05). Participants received compensation in the form of a \euro 10 gift card for their participation. The experiment lasted approximately one hour. Participants spent around 10 minutes to complete an initial phase, consisting of the consent process and a standardized video tutorial, and a median duration of 48 minutes, including self-timed breaks between trials, to perform the task. 

Eligibility criteria were designed to be inclusive, without imposing restrictive measures to ensure a more diverse sample, aiming to enhance the representativeness and generalizability of the study's findings. Most participants (26 out of 34) reported not being regular gamers. All the procedures were approved by the CNR Ethics committee.

\subsection{Experimental task and procedure}

Following the informed consent agreement, participants were asked to see a standardized video tutorial where the task and the experimental design were explained. After ascertaining the task was clear, participants were asked to perform a foraging task in the form of a computer-based video game. To ensure consistency in participant experience, all the data was collected using the same laptop with an external mouse.

\begin{figure} [!htbp]
\centering
\includegraphics[width=1\linewidth]{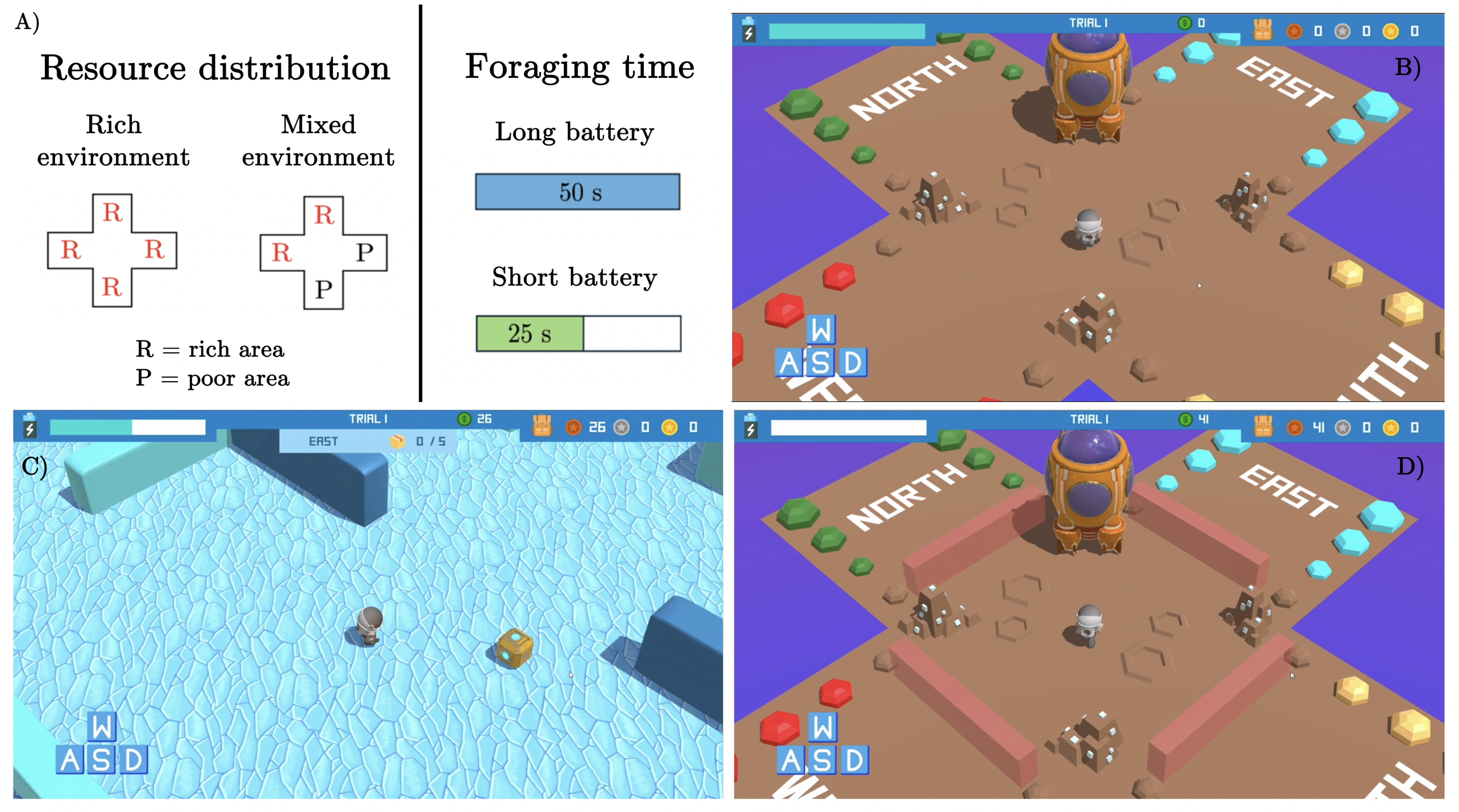}
     \caption {Task design and screenshots from the game illustrating different stages of a trial. (a) Graphical representation of the task design (2x2) showing the two manipulated variables: resource distribution in the environment, either characterized by all areas being rich ("rich environment") or by two rich and two poor areas ("mixed environment"), and time availability, lasting for 50 seconds ("long battery") or 25 seconds ("short battery").
     (b) Trial beginning: the virtual astronaut stands in the middle of the central zone, facing the spaceship. The four distinct areas are visible, the temporal bar at the top left of the screen is fully charged, and the coin counter at the top right is empty. Note that only the leftmost (bronze-colored) coins are used in this experiment, the other two types of coins are not. (c)  Within areas exploration and foraging ("East" area in shown): the temporal bar decreased with time, reflecting the time spent by the astronaut in the area (the temporal bar did not decrease in the central zone).  As shown in the top center of the screen, up to five boxes can be opened per area to collect coins. (d) Trial end: The temporal bar is now empty, and the coin counter displays the total number of coins collected during the trial. The previously accessible areas are not accessible anymore. Only the spaceship can be accessed, enabling players to activate the "reload" function and begin a new trial.}
     \label{fig:screenshots}
 \end{figure}

In the experiment, created with Unity (Unity Technologies, San Francisco, US), participants controlled a virtual astronaut from a third-person perspective, using arrow keys (up, left, down, right) or the letters WASD to navigate. 
The virtual environment consisted of a central zone and four adjacent areas, labeled North, South, East, and West, where "treasure boxes" (henceforth, boxes) were located (Figure ~\ref{fig:screenshots}). Each box contained a variable number of coins and participants' goal was to maximize the number of coins collected during the experiment.

We employed a 2x2 within-subject design to investigate the impact of resource distribution in the environment ("rich environment" vs. "mixed environment") and time availability ("long battery" vs. "short battery") on participants' foraging strategies (Figure~\ref{fig:screenshots}.a). The experiment comprised a total of 2 blocks, each consisting of 20 trials, for a total of 40 trials. To mitigate potential biases, participants were divided into two groups. One group experienced the 20 "long battery" trials in the initial block followed by the 20 "short battery" trials in the subsequent block, while the other group experienced the reverse order. While battery length was manipulated between blocks, the environmental richness was varied within blocks, with each trial being randomly assigned to either a "rich " or a "mixed " environment condition. This implies that participants had to check the sensor at the beginning of each trial, to know whether the environment was rich or mixed (but remind that the sensor did not tell participants which areas was rich or poor). To familiarize participants with the task, we administered 5 training trials before the starting of the first block. Additionally, 1 inter-block training trial was introduced to make participants accustomed to the variation in battery length between blocks.

Participants could either deal with experimental conditions were all four areas are rich (referred to as the "rich environment" condition) or experiencing a combination of two rich and two poor areas (referred to as the "mixed environment" condition). Note that in the "mixed environment" conditions poor and rich areas are randomly distributed across the four areas. The distinction between rich and poor areas did not lie in the number of boxes contained in it, which is always five, but in the number of coins within them.  In rich areas the amount of coins per box decreased linearly from 8 to 4 coins among the five boxes (i.e., the first opened box always delivered 8 coins, the second opened box 7 coins, etc.). Rather, in poor areas, the number of coins per box decreased linearly from 5 to 1 coin (i.e., the first opened box always delivered 5 coins, the second opened box 4 coins, etc.). Participants were informed about the presence of a spaceship in the central zone, which worked as
a sensor (Figure~\ref{fig:screenshots}.b; sensor not shown in the figure). The sensor provided participants with free information about the overall richness of the environment in the current trial, through colored (yellow) bars. Specifically, four long bars indicated the presence of a rich environment, in which all four areas are rich, whereas two long bars and two short bars indicated a mixed environment, in which two areas are rich and two are poor. Since all participants consulted the sensor at the beginning of each trial, we assume that they always knew whether the environment was currently "rich" or "mixed". Note however that the sensor did not provide information about the specific areas designated as rich or poor for each trial -- and rich and poor areas were allocated randomly at each trial. The sole method to ascertain the richness of an area involved actively exploring it and collecting boxes, which subsequently displayed the coin amounts contained in it (Figure~\ref{fig:screenshots}.c).

In addition to manipulating the distribution of resources in the environment, the experiment also involved the manipulation of foraging time available to participants. Participants experienced either trials with a prolonged foraging duration of 50 seconds (referred to as the "long battery" condition) or trials with a shorter duration of 25 seconds (referred to as the "short battery" condition) during trials. Note that the time required to travel between boxes within the same area, or to travel from one area to another, was kept consistent across all experimental conditions (rich/poor areas, long/short battery conditions). The available time was accessible by observing the temporal bar in the top left of the screen; it gradually shortened as time elapsed (Figure~\ref{fig:screenshots}.c,d). Time depletion paused when the astronaut was in the central zone but moved on while in the four surrounding areas. Participants could freely exit any area, returning to the central zone by moving towards any boundary of the area they were foraging within. The time cost in this experiment was implicit in the movement required for harvesting boxes and traveling between areas. There was no explicit time delay or cost added for these actions. Participants could revisit any area within a trial; however, boxes that had already been collected did not refill and remained empty.

At the end of each trial, when the foraging time ended, the virtual astronaut was automatically transported to the central zone where it can only be moved towards the spaceship to activate the "reload" function. By doing so, the coin counts resets to zero, the temporal bar replenishes, and the subsequent trial starts, with a novel (random) assignment of rich or poor areas (Figure~\ref{fig:screenshots}.d).

Note that this experiment provides a rich arena to study key features of foraging scenarios. First, it required virtually navigating to locate and then open the boxes, which were spread in pseudo-random locations in the four areas. Each area contained ten predefined treasure box locations, which remained the same for all the study. However, for each trial, only five (randomly selected) out of these ten locations contained boxes. Note that once all five boxes have been opened in an area, that area becomes empty. Second, it required making stay-or-leave decisions between continuing foraging in the same area versus changing area. In turn, changing area was associated with both motor costs, as it took almost twice as long to change area and collect a box in the new area (average time: 6.10 sec) compared to continuing foraging in the same one (average time: 3.41 sec), and cognitive costs, i.e. keeping track of which areas have been explored and which haven't. Third, in mixed environment, it required exploring to identify which areas were rich and which were poor. Furthermore, it required adapting foraging strategies not just to reward distribution but also to time limitations, as signaled by battery length. Finally, and importantly, this experiment required managing multiple sources of uncertainty, some of which can be reduced over time (e.g., how many boxes and coins they can collect on average in the "long" and "short battery" conditions, how much time it takes on average to reach the next box in the same area or after changing area, where are boxes located and how to reach them most efficiently) and some that cannot (e.g., which areas are rich or poor in a specific trial). Therefore, this experiment permits studying learning dynamics during foraging and asking whether participants' performance increases over time -- perhaps approximating an optimal forager -- as they reduce uncertainty about the task.


\subsection{Analysis methods}

In order to test the influence of resource distribution and time availability on participants’ foraging strategies, we employed linear mixed effects modeling, implemented by the \texttt{Lmer} class from the \texttt{pymer4} library (version 0.8.2) in Python 3.12.3, using a significance level of $p < .05$. 
The dependent variables considered in the analysis are six: the average number of coins gained per trial (i.e. total coins), the number of boxes collected in the first area (i.e., boxes collected in first area), the number of boxes opened per trial (i.e., box collected), the average number of coins per box gained per trial (i.e., average coins per box), the number of visited areas entered only once (i.e., number of visited areas), and the time to move from one box to another, within the same area (i.e., average box collection time). 

The fixed effects in our model included battery length, environment richness, and first area richness, depending on the analysis. The random effects consisted of a random intercept for each subject.
After fitting the LMMs, we conducted pairwise contrasts on the estimated marginal means to identify significant differences between the conditions. To ensure robustness and adjust for multiple comparisons, we used the Tukey method, which is a post-hoc analysis technique that controls the family-wise error rate.
In Section \ref{sec:Learning} and \ref{sec:Optimality} the significant differences between conditions were assessed by means of the Mann-Whitney test.

\subsection{Optimality of foraging strategies}
\label{sec:DecisionTree}

\begin{figure}[!htbp]
\centering
\includegraphics[width=0.8\linewidth]{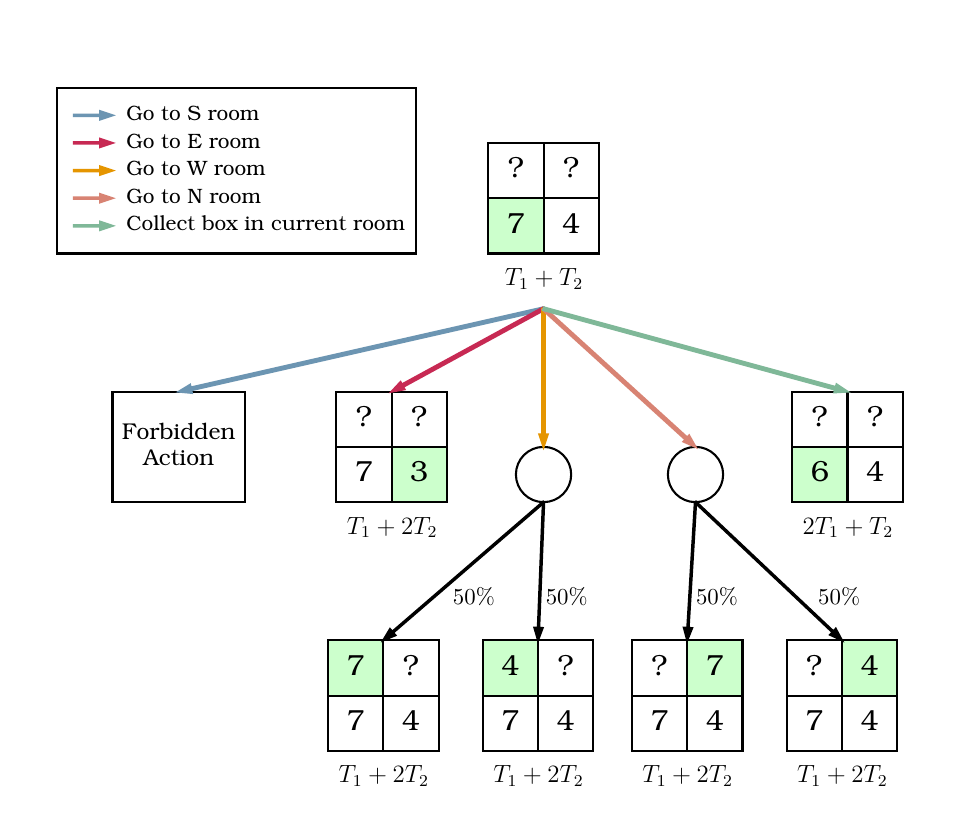}
     \caption {Example branch of the stochastic decision tree for a mixed environment. The agent begins in the $S$ (south) area which is a rich area, knowing that the next box will contain $7$ coins. Moreover, the $E$ (east) area has already been visited and the agent knows that it is a poor area and that the next box will contain $4$ coins. Moving actions are color-coded as follows: blue/red/yellow/peach arrows for moving to the south/east/west/north areas respectively, and the green arrow to remain in the current area and collect a box. The area that is currently occupied by the agent is shown in green. The bottom left cell corresponds to the South (S) area, the bottom right to East (E), the top left to West (W), and the top right to North (N). The action that moves to the south area is forbidden as the agent is currently in the south area. Moving to the east area and staying in the south area and collecting the next box are both deterministic actions, as the outcome can be perfectly predicted by the agent. The remaining two actions, moving to the west and north area, both lead to a stochastic node. Each node has a $50\%$ probability to reveal that the chosen area is rich and $50\%$ probability to reveal that it is poor. The total time associated with each node reflects the cumulative time spent: initially, one box was collected in the East area ($T_1$), followed by a move to the South area and collection of one box there ($T_2$), summing to $T_1 + T_2$; each subsequent action adds either $T_1$ (staying and collecting another box) or $T_2$ (moving to a new area), yielding updated totals such as $2 T_1 + T_2$ or $T_1 + 2 T_2$.}
     \label{fig:DecisionTree}
\end{figure}

To evaluate participants' performance during the foraging task, we developed a model to simulate the decisions of an optimal agent. Note that we could not directly use the Marginal Value Theorem from optimal foraging \citep{charnov1976optimal}, since it considers an infinite time horizon scenario when the goal is to maximize the gain rate, whereas in our experiment time is constrained and an optimal strategy should maximize the total resources collected. Note also that the only source of uncertainty of the model is where rich areas are located in the "mixed" environment. Unlike human participants, the model has no uncertainty about (for example) the time to travel from one box to the other within the same area or across areas. It therefore provides an upper bound to participants' performance.

We constructed a stochastic decision tree to represent all possible trajectories of the optimal agent. The tree's states were defined by three variables: the current trial time, the known state of the environment, and the agent's current area. The environment's state is represented as a tuple of four values, indicating the resource quantities in each area, as known by the agent. For example, $(7,4,?,?)$ describes an environment where the next box in the first area will contain 7 coins, the next box in the second area will contain 4 coins, and the resource quantities in the other two areas are unknown. The possible tuples depend on the environment's richness and the agent's previous trajectory. If the environment is known to be rich, the tuple $(7,?,?,?)$ is impossible, as the agent would know all areas are rich, resulting in a state like $(7,8,8,8)$. Similarly, if the environment is mixed, the tuple $(7,2,6,?)$ is impossible since, after discovering two rich areas and one poor area, the agent would infer that the remaining area must be poor, resulting in a state like $(7,2,6,5)$.

The optimal strategy depends on both the specific trial conditions and individual characteristics of each participant. To compare participants and optimal agents side-by-side, we created four decision trees for each participant, one for each condition (i.e., combination of battery length and environment richness). Additionally, we endowed the models with two participant-specific parameters: the collection time $T_1$, which is the time required to collect a box within a area, and the travel time $T_2$, which is the time needed to leave the current area, move to another area, and collect the first box there. We estimated both times for each participant from the data collected throughout the entire experiment.

The optimal agent can choose between two types of actions in each state: staying in the current area to collect another box, consuming $T_1$ units of time, or moving to a different area to collect a box, consuming $T_2$ units of time. The latter action is further divided into four specific actions, one for each area. Depending on the particular state, some actions are prohibited. For instance, the agent cannot move to the north area if already there, nor can the agent collect a box in the current area if it is empty. An example branch of the decision tree for a mixed richness environment is shown in Figure~\ref{fig:DecisionTree}.

We identified the optimal strategy in each decision tree, as the one that maximizes the total number of coins collected. We assigned a value to each action and state in the decision tree. The value of each state is determined by the maximum value of the actions available from that state. For terminal nodes, where the remaining time is insufficient for further actions, the value is simply the total amount of coins collected. The value of each action is calculated as the immediate reward gained (the number of coins collected) plus the value of the state reached after performing the action. For stochastic transitions, where multiple states could be reached from a given action, the action value is computed as the expected value of all possible state values, weighted by the probability of each transition.

Each decision tree was evaluated through a recursive process, starting from the root node (where no areas have been explored) and proceeding to the terminal nodes. To construct the optimal decision path, we selected the best (highest value) action at each state and evaluated the statistics of interest (total coins collected, number of boxes opened, number of areas visited, etc.) along that path. Due to the presence of stochastic transitions, the optimal strategy does not yield a single optimal trajectory. Therefore, to accurately estimate the statistics of interest, we simulated $10^5$ trajectories for each subject and condition and averaged the results. This procedure permits comparing side-by-side participants' performance with the performance of an optimal agent that solves the same decision trees.

\section{Results}

\subsection{Influence of resource distribution and foraging time on participants’ foraging strategies} 
\label{sec:results_all_trials}

The analyses described in this section show the effects of our manipulations (i.e., battery length and environmental richness, resulting in a 2x2 analysis) on our six dependent variables (i.e., the total coins gained per trial, the boxes opened in the first area, the boxes collected per trial, the average coins gained per box, the number of areas visited, and the average box collection time).

\begin{figure}[!htbp]
\centering
\includegraphics[width=0.9\linewidth]{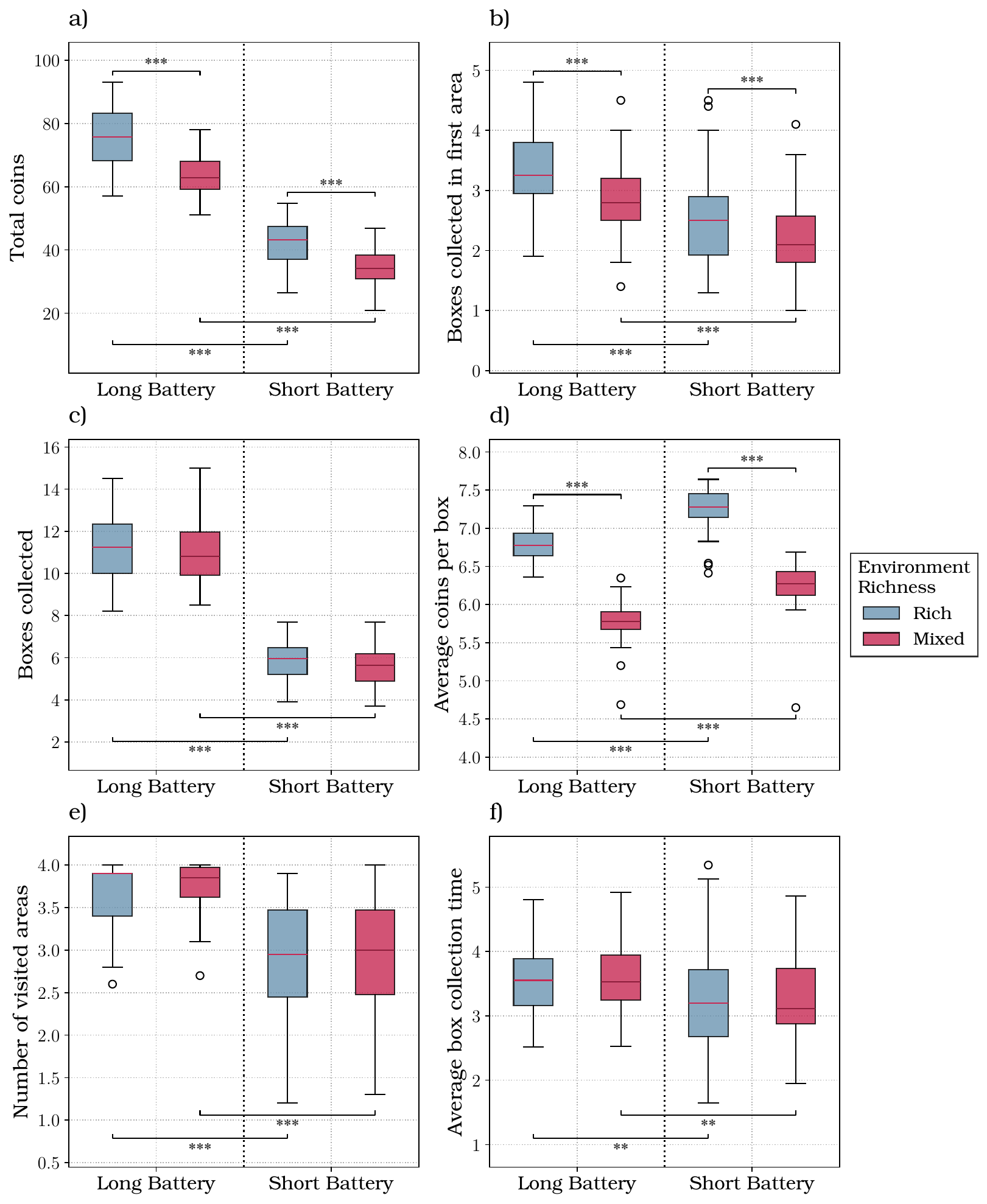}
     \caption {Results of the foraging experiment. The box plots show the six dependent variables considered in this study: (a) the total coins; (b) the boxes collected in the first area; (c) the boxes collected; (d) the average coins per box; (e) the number of visited areas; (f) the average box collection time. The plots consider the two levels of each independent variables (i.e., battery length, either long or short, and richness of the environment, either rich or mixed).}
     \label{fig:Boxplot_Battery_Environment_Vertical}
\end{figure}

We first considered the total coins gained per trial, which provides an aggregate measure of participants' performance (Figure~\ref{fig:Boxplot_Battery_Environment_Vertical}.a; Table~\ref{tab:coinsgainedpertrial_2x2}). The linear mixed modeling analysis unveiled a significant effect of both our manipulations and their interaction on this variable. The battery length significantly influenced the amount of coins collected by participants (Coeff.: 28.356; SE: 0.673; t: 42.161; \(P\)-value: $< 0.001$; 2.5, 97.5 ci: [27.038, 29.674]). Specifically, the longer the battery, the more the collected coins. The richness of the environment also showed a main effect on the dependent variable (Coeff.: 7.241; SE: 0.673; t: 10.767; \(P\)-value: $< 0.001$; 2.5, 97.5 ci: [5.923, 8.559]). Indeed, in a rich environment trial, the amount of coins available in the environment is higher than in mixed environment scenarios. Furthermore, the interaction between the two variables revealed to be significant (Coeff.: 5.600; SE: 0.951; t: 5.888; \(P\)-value: $< 0.001$; 2.5, 97.5 ci: [3.736, 7.464]). Specifically, participants collected a significantly greater amount of coins when experiencing long battery and rich environment trials' combinations. 

We next considered the number of boxes collected in the first area, which is a more fine-grained measure of the patch-leaving strategy adopted by participants (Figure~\ref{fig:Boxplot_Battery_Environment_Vertical}.b; Table~\ref{tab:boxopenedinthefirstarea_2x2}). The linear mixed modeling analysis unveiled a significant effect of battery length (Coeff.: 0.600; SE: 0.081; t: 7.431; \(P\)-value: $< 0.001$; 2.5, 97.5 ci: [0.442, 0.758]) and richness of the environment (Coeff.: 0.318; SE: 0.081; t: 3.934; \(P\)-value: $< 0.001$; 2.5, 97.5 ci: [0.159, 0.476]). When the environment is mixed and the battery length is short, participants forage for a significantly lower amount of boxes in the first area. As a follow-up analysis, we compared the number of boxes collected in the first area, by restricting the analysis only to the cases in which it was rich (which is always true in the rich environment, but only approximately half of the times in the mixed environment). This follow-up analysis shows that participants collected more boxes in the mixed environment compared to the rich environment, for both long and short batteries (Figure~\ref{fig:boxplotfirstarearich_boxesfirst} and Table~\ref{tab:boxopenedfirstarearich}).

We also analyzed the number of boxes collected per trial (Figure~\ref{fig:Boxplot_Battery_Environment_Vertical}.c; Table~\ref{tab:boxopenedpertrial_2x2}). The mixed model analysis indicated that the battery length (Coeff.: 5.432; SE: 0.113; t: 48.045; \(P\)-value: $< 0.001$; 2.5, 97.5 ci: [5.211, 5.654]) and the richness of the environment (Coeff.: 0.244; SE: 0.113; t: 2.159; \(P\)-value: 0.031; 2.5, 97.5 ci: [0.023, 0.466]) significantly affect the dependent variable. However, while the environment richness effect was significant as a main effect in the linear mixed model, pairwise post-hoc contrasts between rich and mixed environments did not reach significance under either battery condition. As a follow-up analysis, we compared the number of coins collected before leaving rich areas (Figure \ref{fig:CoinsForArea}.a) and poor areas (Figure \ref{fig:CoinsForArea}.b). We found significant differences between the number of coins collected under the different conditions, ruling out the possibility that participants followed a fixed, threshold-driven strategy to decide when to leave areas.

Furthermore, we explored the effect of our manipulations on the average number of coins collected per box (Figure~\ref{fig:Boxplot_Battery_Environment_Vertical}.d; Table~\ref{tab:coinsgainedpertrialmean_2x2}). The results from the linear mixed model unveiled a significant and negative influence of battery length (Coeff.: -0.454; SE: 0.032; t: -14.167; \(P\)-value: $< 0.001$; 2.5, 97.5 ci: [-0.391, 0.851]) on this average. This suggests that the less the available foraging time, the more participants try to maximize their coin gains, which increases this average number. The environmental richness significantly affects the average number of coins collected per box, too (Coeff.: 0.999; SE: 0.032; t: 31.182; \(P\)-value: $< 0.001$; 2.5, 97.5 ci: [0.936, 1.061]). Foraging in a rich environment increases the average number of coins collected per box.

We also examined the average number of unique areas visited per trial (namely, each area is counted only once regardless of how many times it is visited during the trial)(Figure~\ref{fig:Boxplot_Battery_Environment_Vertical}.e; Table~\ref{tab:visited_areas_2x2}). The linear mixed modeling analysis showed a significant influence of the battery length (Coeff.: 0.765; SE: 0.044; t: 17.342; \(P\)-value: $< 0.001$; 2.5, 97.5 ci: [0.678, 0.851]). The longer the available time, the greater the number of areas participants visited per trial.

Finally, we examined the impact of our manipulations on the time required to collect each box, computed as the mean time between consecutive boxes within the same area for each trial (Figure~\ref{fig:Boxplot_Battery_Environment_Vertical}.f; Table~\ref{tab:averageboxcollectiontime_2x2}). The battery length showed a significant impact on this variable (Coeff.: 0.263; SE: 0.073; t: 3.597; \(P\)-value: $< 0.001$; 2.5, 97.5 ci: [0.120, 0.406]):with a longer duration of the battery, participants' navigation time between boxes was slower. 

\subsection{Influence of resource distribution and foraging time on participants’ foraging strategies, in the mixed environment}
\label{sec:mixed_env_analysis}

\begin{figure}[!htbp]
\centering
\includegraphics[width=0.9\linewidth]{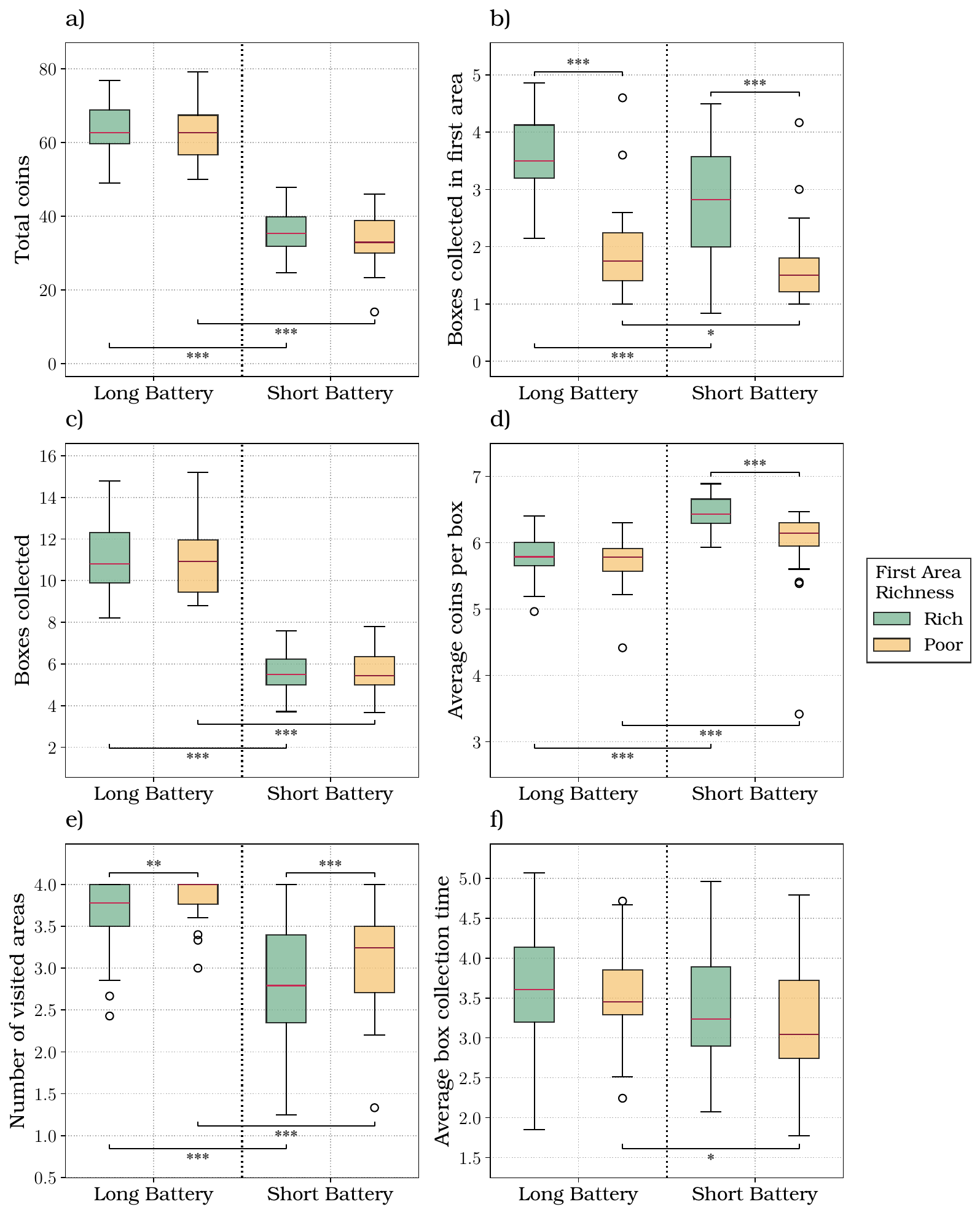}
     \caption{Results of the foraging experiment, when considering the mixed environment only. The box plots show the six dependent variables considered in this study: (a) the total coins; (b) the boxes collected in the first area; (c) the boxes collected; (d) the average coins per box; (e) the number of visited areas; (f) the average box collection time. The plots consider the two levels of our directly manipulated independent variable (i.e., battery length, either long or short) and the two levels of our indirectly manipulated variable (i.e., richness of the first area visited in the mixed environment, either rich or poor).}
     \label{fig:Boxplot_Battery_FirstArea_Vertical}
\end{figure}

The 2x2 analysis reported in Section \ref{sec:results_all_trials} did not consider that when foraging in a mixed environment, participants might select as the first area either a rich or a poor area, with equal probability (50\%) -- and this initial (random) choice might change their strategy. For this, we conducted another linear mixed model analysis, by taking into account the richness of the first area visited by participants, in the mixed environment trials (Figure~\ref{fig:Boxplot_Battery_FirstArea_Vertical}). 

We first considered the total coins gained per trial (Figure~\ref{fig:Boxplot_Battery_FirstArea_Vertical}.a; Table~\ref{coinsgainedpertrial_2x2x2}). The linear mixed modeling analysis revealed that the battery length (Coeff.: 29.008; SE: 0.836; t: 34.716; \(P\)-value: $< 0.001$; 2.5, 97.5 ci: [27.370, 30.646]) and the richness of the first area significantly affected the total amount of coins gained per trial (Coeff.: 1.931; SE: 0.822; t: 2.349; \(P\)-value: 0.019; 2.5, 97.5 ci: [0.320, 3.542]), with more coins gained in trials in which the battery was long and the first visited area was rich.

We next considered the number of boxes opened in the initial area (Figure~\ref{fig:Boxplot_Battery_FirstArea_Vertical}.b; Table~\ref{tab:boxopenedfirstarea_2x2x2}). The linear mixed modeling analysis revealed a significant influence of battery length on this variable (Coeff.: 0.290; SE: 0.100; t: 2.893; \(P\)-value: 0.004; 2.5, 97.5 ci: [0.093, 0.486]). Participants opened a significantly higher amount of boxes in the first area during long battery trials. Moreover, a significant main effect of the richness of the first area (Coeff.: 1.205; SE: 0.098; t: 12.239; \(P\)-value: $< 0.001$; 2.5, 97.5 ci: [1.012, 1.398]) emerged from the analysis, with more boxes opened in trials in which the first visited area was rich. Notably, the interaction between the battery length and the richness of the initial area (Coeff.: 0.479; SE: 0.139; t: 3.451; \(P\)-value: 0.001; 2.5, 97.5 ci: [0.207, 0.751]) significantly affected the number of boxes opened in it. Participants opened a greater number of boxes in the initial area when it was rich and their battery duration was longer.

Focusing on the number of boxes opened per trial (Figure~\ref{fig:Boxplot_Battery_FirstArea_Vertical}.c; Table~\ref{tab:boxopenedpertrial_2x2x2}), the linear mixed modeling analysis revealed a significant effect of battery length (Coeff.: 5.404; SE: 0.159; t: 33.997; \(P\)-value: $< 0.001$; 2.5, 97.5 ci: [5.092, 5.715]), implying that participants engaging in trials with longer battery duration collected a significantly greater number of boxes.

We next focused on the average number of coins collected per box (Figure~\ref{fig:Boxplot_Battery_FirstArea_Vertical}.d; Table~\ref{tab:coinsgainedpertrialmean_2x2x2}). The analysis revealed a significant negative impact of the battery length on this variable (Coeff.: -0.263; SE: 0.053; t: -4.953; \(P\)-value: $< 0.001$; 2.5, 97.5 ci: [-0.367, -0.159]). In trials with long battery length, participants engaged in more extensive foraging making use of the extra time to even search for less rewarding boxes. As a consequence of that, this behavior resulted in a reduced average number of coins gained per box. Moreover, the richness of the first area significantly influenced this dependent variable (Coeff.: 0.455; SE: 0.052; t: 8.724; \(P\)-value: $< 0.001$; 2.5, 97.5 ci: [0.353, 0.557]). Specifically, the average number of coins collected per box was significantly greater when the first visited area was rich. Further analysis revealed a significant, albeit negative, effect of the interaction between the battery length and the richness of the initial area (Coeff.: -0.385; SE: -0.241; t: -5.242; \(P\)-value: $< 0.001$; 2.5, 97.5 ci: [-0.530, -0.241]) on this average. In conditions when the first area was poor and the battery length was short, participants collected a greater amount of coins per box. 

We next examined the number of areas participants visited per trial across different conditions (Figure~\ref{fig:Boxplot_Battery_FirstArea_Vertical}.e; Table~\ref{tab:visited_areas_2x2x2}). The linear mixed modeling analysis unveiled a significant influence of battery length on the number of visited areas (Coeff.: 0.718; SE: 0.064; t: 11.199; \(P\)-value: $< 0.001$; 2.5, 97.5 ci: [0.592, 0.844]), indicating that participants explored a significant greater number of areas with increased available time. Additionally, the analyses revealed a negative main effect of the richness of the first area on the number of visited areas (Coeff.: -0.326; SE: 0.063; t: -5.171; \(P\)-value: $< 0.001$; 2.5, 97.5 ci: [-0.450, -0.202]). This result suggests that when their first visited area was rich, participants foraged in the same area for longer, then visiting fewer areas overall.

Finally, we examined the effect on the average box collection time (Figure~\ref{fig:Boxplot_Battery_FirstArea_Vertical}.f; Table~\ref{tab:averageboxcollectiontime_2x2x2}). The analysis revealed a significant influence of battery length on foraging time between boxes (Coeff.: 0.291; SE: 0.109; t: 2.679; \(P\)-value: 0.008; 2.5, 97.5 ci: [0.078, 0.505]), indicated that with more the available time, participants were slower in navigating from one box to another.


\subsection{Learning dynamics during the experiment}
\label{sec:Learning}

Our foraging experiment requires (virtual) navigation in a large environment without full visibility and with multiple sources of uncertainty about resource locations and distributions. Hence, during the experiment, participants can learn these variables and potentially improve their performance. 

\begin{figure}[!htbp]
\centering
\includegraphics[width=\linewidth]{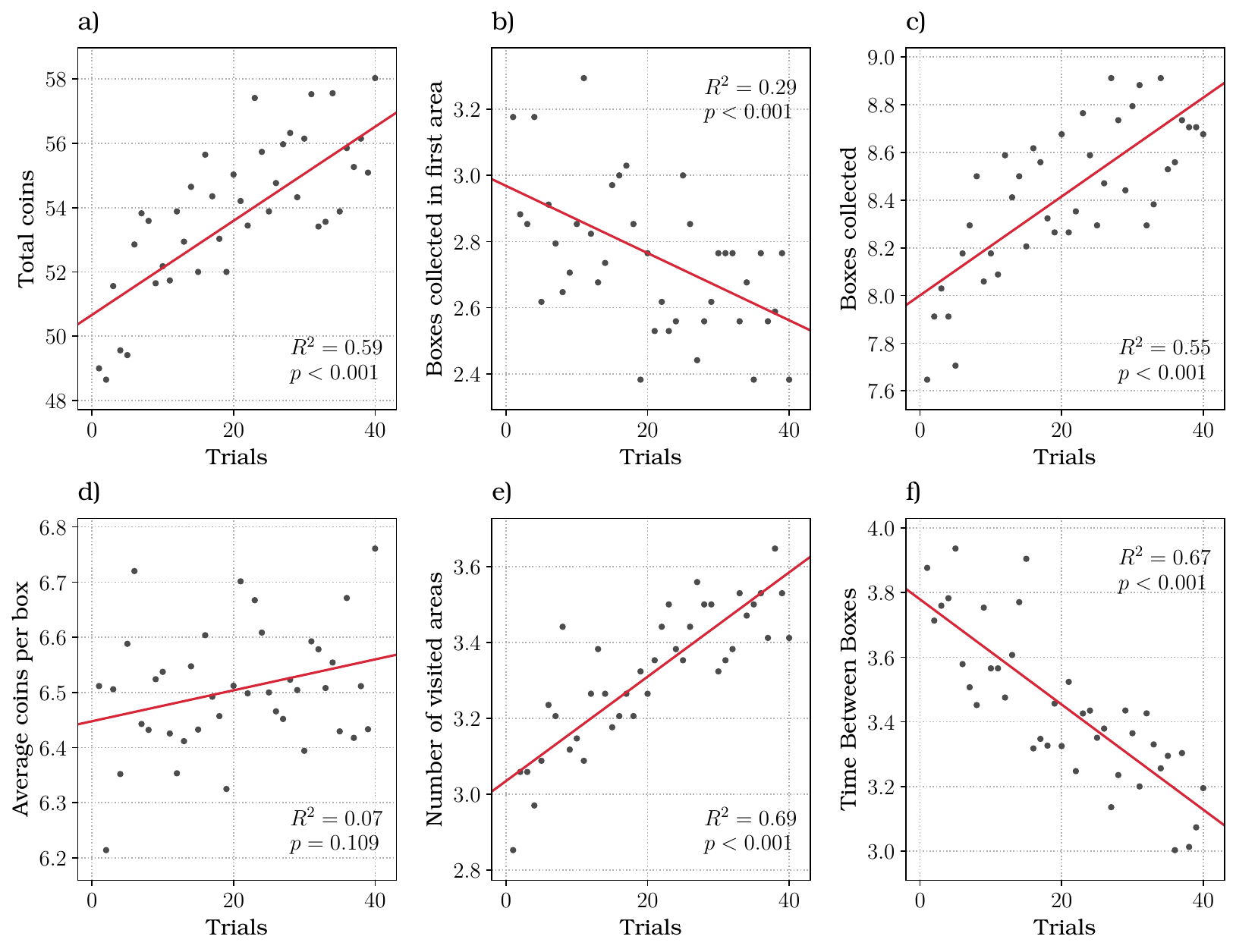}
     \caption {Participants' performance across the 40 experimental trials for our six dependent variables: (a) the total coins; (b) the boxes collected in the first area; (c) the boxes collected; (d) the average coins per box; (e) the number of visited areas; (f) the average box collection time. Each dot represents the mean value computed from the data of our 34 participants for every trial, with each of the 40 trials being represented by a distinct dot. The red line represents the linear regression fit to the data. The coefficient of determination $R^2$ and the \(P\)-value are reported.}
     \label{fig:learning}
\end{figure}

To investigate these learning dynamics, we tested whether our six dependent variables change throughout the course of the experiment, by computing the average of participants data for each trial. It is important to remark that trial presentation is randomized for each participant, hence the average trial data effectively averages across all the experimental conditions. Figure~\ref{fig:learning} illustrates participants' performance across trials, for each of our six dependent variables. Each dot in the figure represents the mean value computed from the data of our 34 participants for every trial, with each of the 40 trials being represented by a distinct dot. Looking at the graphs, the distribution of dots changes across trials, indicating a trend that can be attributed to learning. This suggests that participants' performance evolves over time as they become more familiar with various aspects of the tasks, such as how to navigate, the plausible location of boxes, coin distributions for rich and poor environments, etc.

The red line represents the linear regression fit to the data. Focusing on the total number of coins (Figure~\ref{fig:learning}.a), the coefficient of determination $R^2 = 0.59$ indicates that the $59\%$ of the variance in the total number of coins can be explained by the independent variables and that the observed relationship is statistically significant (\(P\)-value: $< 0.001$). This result suggest that as people become more familiar with the task, they collect more coins. Further analysis on the amount of boxes collected in the first area suggests that the increased performance was at least partially determined by a change in patch-leaving strategy across trials (Figure~\ref{fig:learning}.b). Indeed, the scatter plot shows a significant ($R^2 = 0.29$; \(P\)-value: $< 0.001$) and negative trend across trials indicating that people tended to overharvest less, the more they became familiar with the task. Moreover, as participants became more familiar with the task, they collected a significantly greater number of boxes per trial ($R^2 = 0.55$; \(P\)-value: $< 0.001$) (Figure~\ref{fig:learning}.c) and visited more areas ($R^2 = 0.69$; \(P\)-value: $< 0.001$) (Figure~\ref{fig:learning}.e). Training may account for these outcomes, as demonstrated by the fact that participants' became faster at moving from one box to another during the course of the experiment ($R^2 = 0.67$; \(P\)-value: $< 0.001$) (Figure~\ref{fig:learning}.f). Additionally, participants' average \emph{exit time} (i.e., the time elapsed between collecting the last box in an area and exiting that area) decreases over time (R² = 0.19, p = 0.005), suggesting that participants became more efficient at navigating and making stay-or-leave decisions during the experiment (Figure \ref{fig:learningExit}). Finally, the average coins gained per box did not show a significant variation across trials ($R^2 = 0.07$; \(P\)-value: $< 0.109$) (Figure~\ref{fig:learning}.d). This is likely due to a combination of collecting more coins, which increases the average, and opening more boxes, which decreases it.


More qualitatively, we observed that while navigation strategies varied across participants, they remained stable over time. Some participants consistently followed a "clockwise" strategy, always starting from the same area—possibly to reduce memory load (see Figure \ref{fig:navigationalpatterns}.a for an example). Others, however, selected a different starting area in each trial (see Figure \ref{fig:navigationalpatterns}.b for an example).

\begin{figure}[!htbp]
\centering
\includegraphics[width=\linewidth]{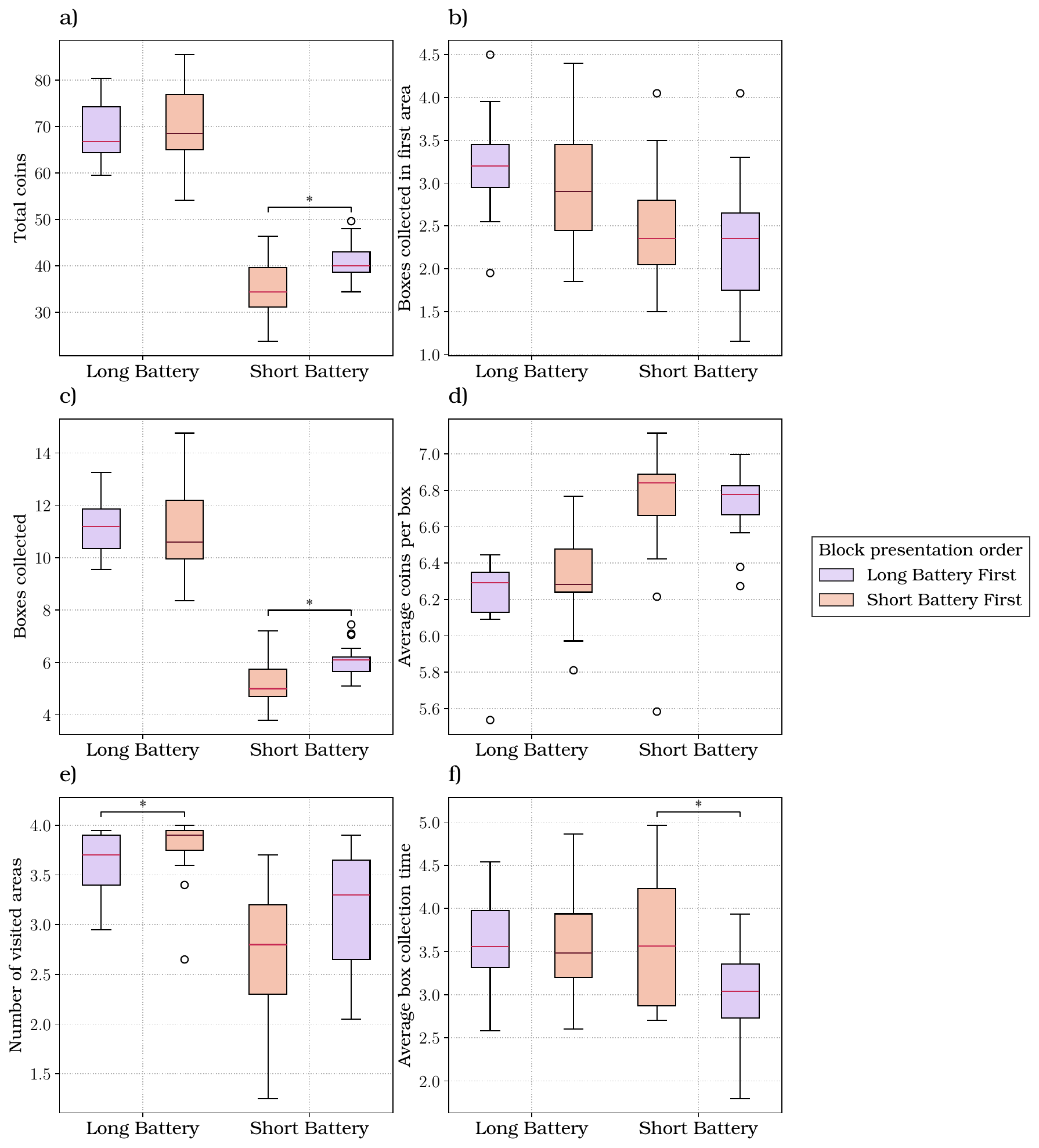}
     \caption {Group differences between participants who experienced the "long battery" or the "short battery" blocks first. From (a) to (f), the box plots represent: (a) the total coins; (b) the boxes collected in the first area; (c) the boxes collected; (d) the average coins per box; (e) the number of visited areas; (f) the average box collection time.}
     \label{fig:LearningDifferentBlocks}
\end{figure}

We also considered that in our experiment, participants were divided into two groups: one group performed the "long battery" block first, followed by the short battery block, while the other group performed the two blocks in the reverse order. Given that the group that experienced the long battery block first had twice the time for learning compared to the other group, we aimed to investigate whether they performed better in the second (short battery) block. For this, we divided participants' data according to the two groups and we performed non parametric Mann-Whitney U tests on each of our six dependent variables (Figure~\ref{fig:LearningDifferentBlocks}). We found that participants who started the experiment with the long battery block earned more coins (Figure~\ref{fig:LearningDifferentBlocks}.a, \(P\)-value: $< 0.019$), opened more boxes (Figure~\ref{fig:LearningDifferentBlocks}.b, \(P\)-value: $< 0.013$) and moved faster between boxes (Figure~\ref{fig:LearningDifferentBlocks}.f, \(P\)-value: $< 0.025$) in the short battery block, compared to participants who started the experiment with the short battery block. Finally, participants who started the experiment with the short battery block visited significantly more areas in the long battery block (Figure~\ref{fig:LearningDifferentBlocks}.e, \(P\)-value: $< 0.049$).

Finally, we asked whether participants could adapt their behavior not only between trials (as considered in the previous analyses) but also \emph{within trials}. We reasoned that as participants navigated across areas, they could have deliberated more or less depending on their level of uncertainty about resource distribution. To test this idea, we repeated the analysis of participants'\emph{exit time} (illustrated in Figure~\ref{fig:learningExit}), but within trials. In mixed environments where the richness of each area is unknown and must be inferred through experience, we expected longer exit times early in the trial—when participants are still unsure about the overall resource distribution—and shorter times later, when uncertainty has been (at least partially) resolved. Conversely, in rich environments where all areas are known to be rich from the start, exit time should not depend on the area visit order, because no additional information is being gathered across areas.

To test this hypothesis, we ran two separate linear mixed-effects models for rich and mixed environments, with exit time as the dependent variable and area visit order as a fixed-effect regressor. The models included random intercepts per subject. The analysis revealed no significant effect of area visit order in the rich environment ($p = 0.343$), suggesting that exit time remained stable across areas. In contrast, we found a significant negative effect in the mixed environment ($p = 0.01$), indicating that participants exited later areas more quickly than earlier ones (Table~\ref{tab:ExitTime}). These findings support the idea that exit time reflects participants' evolving certainty about the environment within each trial.

\begin{table}[!htbp]
    \centering
    \small
    \begin{tabular}{lcccccc}
        \toprule
        \textbf{Rich Environment} & Coeff. & SE & t & 2.5 ci & 97.5 ci & p-value \\
        \midrule
        Intercept & 2.675 &  0.085 & 31.615 & 2.509 & 2.841 & $\bm{< 0.001}$\\
        Area visit order & -0.030 & 0.032 & -0.948 & -0.093 & 0.032 & 0.343\\
        \midrule
        \textbf{Mixed Environment} \\
        \midrule
        Intercept & 2.726 &  0.067 & 40.572 & 2.594 & 2.858 & $\bm{< 0.001}$\\
        Area visit order & -0.073 & 0.028 & -2.562	 & -0.128 & -0.017 & $\bm{0.01}$\\
        \bottomrule
    \end{tabular}
\caption{Results of the linear mixed-effects models assessing whether exit time (in seconds) changes with the area visit order, separately for rich and mixed environments. A significant negative effect of area visit order is observed only in mixed environments, indicating faster exits from later areas as participants reduce uncertainty about resource distribution.}
\label{tab:ExitTime}
\end{table}

\subsection{Assessing the optimality of participants' foraging strategies}
\label{sec:Optimality}

In this section, we investigate whether participants' performance and strategies align with optimal foraging behavior in our task. For each participant and condition, we used the model described in Section \ref{sec:DecisionTree} to simulate an optimal agent's behavior. The two input parameters for each model, the time required to collect a box in the current area ($T_1$) and the time needed to switch area and then collect a box in the new area ($T_2$), were estimated for each participant, by averaging all instances of the corresponding events, regardless of the condition. The average values of these times across all subjects are $T_1 = (3.41 \pm 0.53)s$ and $T_2 = (6.10 \pm 0.72)s$. A Mann-Whitney test comparing the distributions of $T_1$ and $T_2$ across all subjects revealed that the time to collect boxes after switching areas was significantly longer ($p < 0.001$).

\begin{figure}[!htbp]
\centering
\includegraphics[width=\linewidth]{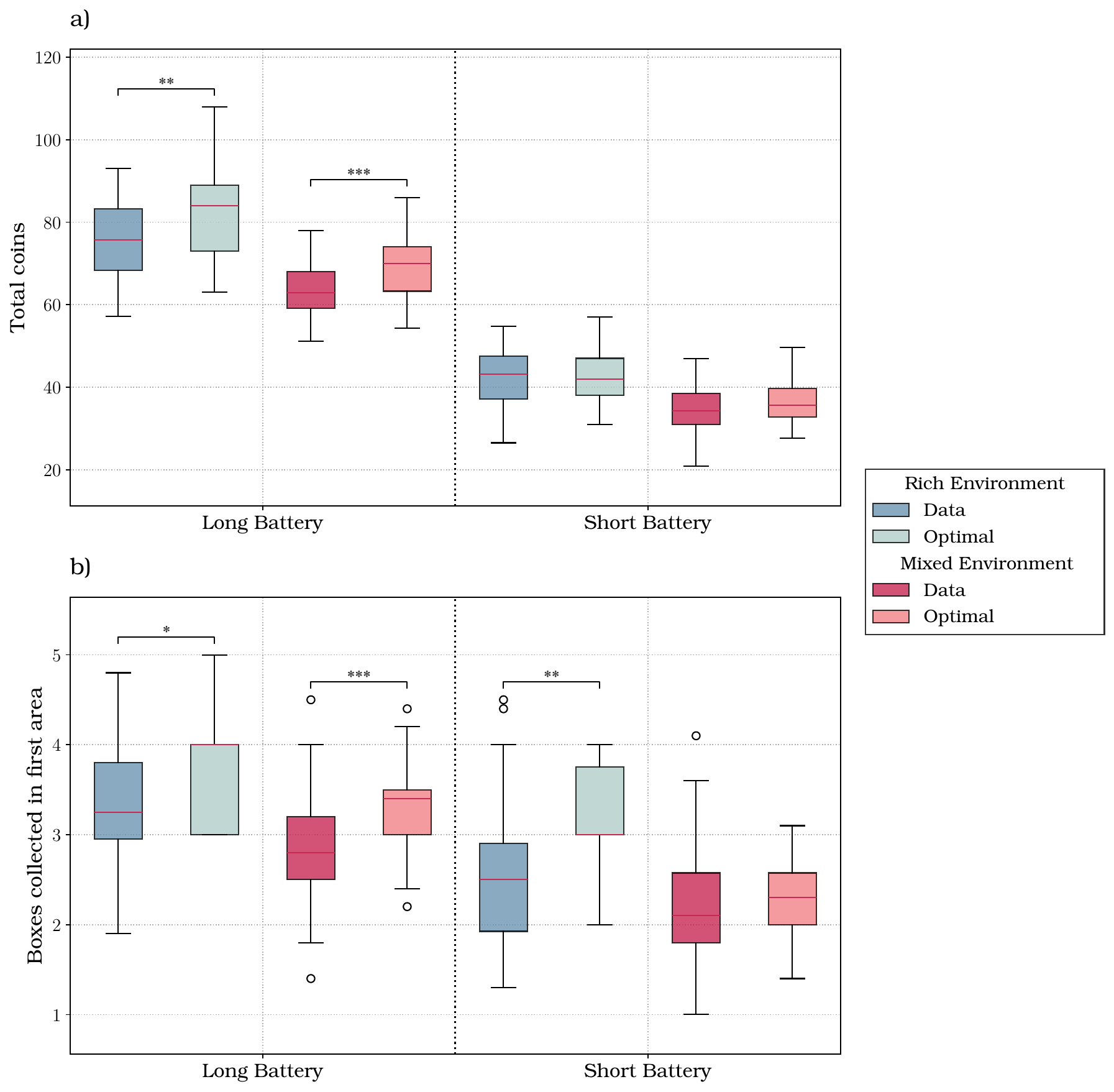}
     \caption {Comparison of the performance of participants and the optimal agent. (a) Total coins collected per trial for long and short battery conditions, in both rich and mixed environments. (b) Number of boxes collected in the first area, for long and short battery conditions. Note that the optimal values for the rich environment conditions always have an integer value, since there is no stochasticity and hence there is only one optimal decision path for the optimal agent. Significant differences between participants' data and the optimal model are indicated (*$p < 0.05$, **$p < 0.01$, ***$p < 0.001$).}
     \label{fig:Optimality}
\end{figure}

In the comparison between participants and optimal agents, we considered two measures: the total number of coins collected, which gives an overall measure of performance; and the number of boxes collected in the first area, which provides a key indication of strategy in patch-leaving problems \citep{charnov1976optimal}. Figure~\ref{fig:Optimality}.a shows the results of the comparison between participants and optimal agents, when considering the average total number of coins collected per trial. In the long battery trials, there is a significant difference between the coins collected by participants and the optimal agent, for both the rich environment condition ($p = 0.006$) and the mixed environment condition ($p < 0.001$). In contrast, for the short battery condition, there are no significant differences between the coins collected by participants and the optimal agent. This implies that participants approximate better the performance of the optimal agent in the short battery condition.

Figure~\ref{fig:Optimality}.b shows the results of the comparison between participants and optimal agents, when considering the number of boxes opened in the first area. For the long battery condition, there is a significant difference between participants and the optimal agent in both the rich ($p = 0.033$) and the mixed ($p < 0.001$) conditions. In both cases, the median value of boxes opened by the optimal agent is greater than the median participant value, indicating that participants tend to underharvest in the long battery condition. Furthermore, there is a significant difference between the boxes opened by participants and the optimal agent in the short battery and rich environment condition ($p = 0.003$), with the optimal agent opening more boxes than participants. 

\begin{figure}[!htbp]
\centering
\includegraphics[width=\linewidth]{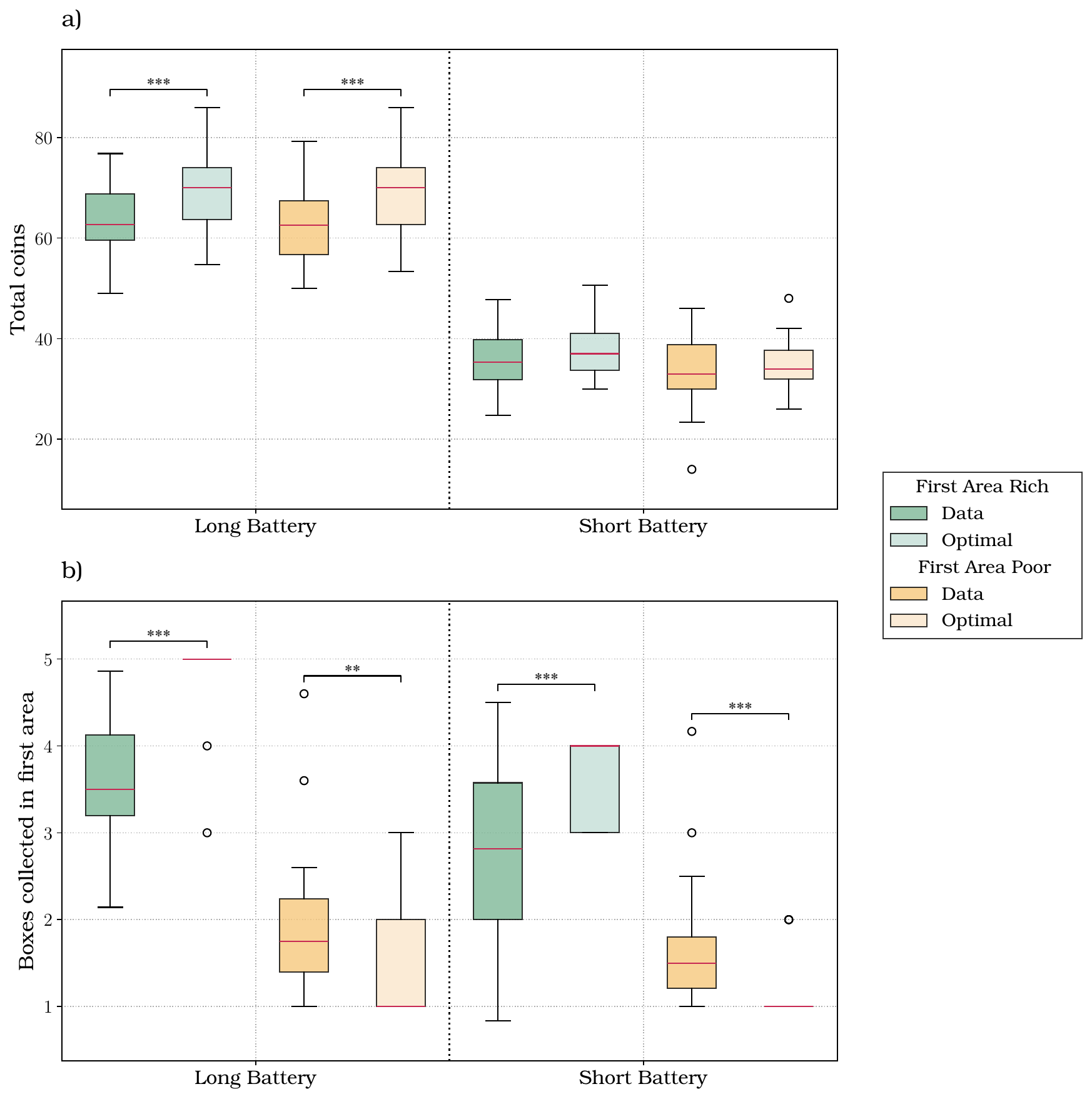}
     \caption {Comparison between participants' results and the optimal model in a mixed-richness environment, with the richness of the first area as a factor. (a) Total coins collected per trial. (b) Boxes collected in the first area. Significant differences between participants' data and the optimal model are indicated (*$p < 0.05$, **$p < 0.01$, ***$p < 0.001$). }
     \label{fig:OptimalityFirstArea}
\end{figure}

We next compared participants and the optimal model in the mixed environment, when splitting data by the richness of the first visited area. Figure~\ref{fig:OptimalityFirstArea}.a shows the results of the comparison between participants and optimal agents, when considering the average total number of coins collected per trial. The results show a significant difference between coins collected by participants and the optimal agent in the long battery condition ($p < 0.001$ for both rich and poor first areas), but not in the short battery condition. Figure~\ref{fig:OptimalityFirstArea}.b shows the results of the comparison between participants and optimal agents, when considering the number of boxes opened in the first area. When the first area was rich, participants collected a significantly smaller number or boxes compared to the optimal agent ($p < 0.001$), indicating a tendency to underharvest in rich areas. Conversely, when the area is poor, participants collected a significantly greater number of boxes compared to the optimal agent, for both long ($p = 0.004$) and short ($p < 0.001$) battery conditions, suggesting a tendency to overharvest in poor areas.

Finally, we considered that (as discussed in Section \ref{sec:Learning}), participants' performance increased over time as they reduced uncertainty about resource locations and distributions, prompting the question of whether it could eventually reach the performance of the optimal model with full knowledge of task statistics. To assess this, we divided the 40 trials into four blocks of 10 trials each and, for each block, we compared the distributions of total coins collected by participants and the optimal model, using four t-tests (Figure~\ref{fig:optimalityTrials}). Our analyses revealed that the optimal model significantly outperforms participants across all the four blocks, with $p < 0.001$ for the first three blocks and $p = 0.005$ for the last block. These results (along with the learning trend shown in Figure \ref{fig:learning}) indicate that participants' performance improves over time, approximating but not reaching the level of the optimal agent.


    


\section{Discussion}

Human and other animals forage for valuable resources, such as food or other goods, in conditions in which the availability of the resources themselves and of foraging time might be limited. However, we still have a limited understanding of the ways the availability of resources and time influence human foraging strategies and whether participants are able to flexibly adapt their strategies to these different conditions as they learn more about the task and its statistics -- and whether they approximate optimal solutions provided by foraging theory \citep{charnov1976optimal}.

To address these challenges, we designed a novel human foraging experiment that manipulates orthogonally the resource distribution and the time availability. Concerning resources distribution, the experiment comprised trials in which all the areas were rich (rich environment) and trials in which two areas were rich and two were poor (mixed environment). Despite a fixed number of "treasure boxes" per area in all the trials (i.e., five boxes), the number of coins varied across rich areas (in which the number of coins per box decreased linearly from 8 to 4 coins) and poor areas (in which the number of coins per box decreased linearly from 5 to 1 coin). Concerning time availability, the experiment comprised two blocks, in which participants had longer (50 seconds) or shorter (25 seconds) foraging time, as signaled by a battery that was always visible to participants. To increase the realism of the situation, we adopted a video-game like scenario that incorporates some aspects of animal foraging, requiring participants not only to make abstract choices between areas and boxes, but also (virtually) navigating between areas, locating boxes and learning their possible locations. This scenario incorporates various sources of uncertainty, about (for example) resources distribution and time required to travel across boxes and areas, and hence various occasions for learning during during the experiment.

We aimed to address three main research questions: first, whether participants' strategies are influenced by resource distributions and time availability, or their combination; second, whether they were able to learn and improve their strategies during the experiment; and third, whether and under which conditions they showed optimal or suboptimal foraging behavior.

Concerning the first question, we found that that participants showed flexibility in foraging behaviour and were sensitive to both resources distribution and time availability. While some results (e.g., more coins collected with longer batteries) are simply the expected effect of our manipulations, others are indicatives of strategy. A key indication of adaptive foraging strategies comes from the analysis of boxes opened in the first area. Participants collected more boxes in the first area when doing so was more favourable, namely, when they were in a rich environment and their battery duration was longer (Figure~\ref{fig:Boxplot_Battery_Environment_Vertical}). Interestingly, when their first visited area was rich, they opened more boxes when the environment was mixed than when it was rich, showing sensitivity to these different conditions (Figure~\ref{fig:boxplotfirstarearich_boxesfirst}). Furthermore, participants modulated their strategy and their effort as a function of both richness of the environment and foraging time. First, participants opened more boxes in rich environments, suggesting that they exploited more and potentially invested more effort with more points to be earned. Second, and more remarkably, participants showed faster navigation times between boxes within the same area when foraging time was shorter. Interestingly, this occurred not only in rich (and less uncertain) environments but also in mixed (and more uncertain) ones. In short battery trials within mixed environments, participants necessarily gathered less information than in long battery trials, leading to greater uncertainty—something that would typically call for more, not less, deliberation and slower movement. The fact that participants were faster rather than slower in short battery trials suggests that limited time prompted greater effort, consistent with previous findings \citep{lewis2004flexible}, without resulting in less efficient foraging \citep{Wu2022}. This aligns with recent work by \citep{pisauro2024neural}, which showed that under deadline pressure, effort can be perceived as a valued means to achieve goals rather than merely a cost to minimize. Taken together, these results show that participants flexibly adapt various aspects of their foraging strategy -- including the choice of when to leave areas and how fast to move -- to resource distribution and time availability. Future studies might assess more fine-grained aspects of adaptive foraging strategy, by manipulating other aspects of the task, such as the uncertainty about rewards and their volatility.

Concerning the second question, we found that during the experiment, participants showed a clear learning trend and an overall increase of performance, as indexed for example by the number of coins collected (Figure \ref{fig:learning}.a). The increase of performance was plausibly due to the reduction of uncertainty about resource locations and distributions -- and to a combination of improved skill and cognitive strategy. First, during the experiment participants learned how to navigate more effectively and what are the most likely locations of boxes, and became faster in locating and opening them (Figure \ref{fig:learning}.f). Second, participants modified their patch leaving strategy across trials. Specifically, they reduced their initial overharvesting and the number of boxes opened in the first area (Figure \ref{fig:learning}.b), as they become more confident that they can effectively locate boxes across all the other areas. Our results suggest a transition between an initial, more exploratory phase in which participants need to learn task statistics to a subsequent phase in which participants develop and exploit a more effective strategy that approximates an optimal forager without uncertainty, but without fully reaching it (see below). It is, however, worth noting that while uncertainty decreases during the task, it cannot be fully eliminated. For example, the locations of the boxes and the time needed to locate and reach them vary from trial to trial, making fixed strategies (e.g., always collecting two boxes before leaving an area) less effective. Furthermore, in "mixed environments," it is only after exploring two areas that a participant can be certain about which areas are rich. Consistent with this, we found that within-trial uncertainty influences participants' behavior and deliberation time. In particular, the observed reduction in exit time across visited areas in mixed environments suggests that participants deliberated more when uncertainty was higher and accelerated their decisions once they had gathered enough information to reduce ambiguity about area richness. Notably, this pattern was absent in rich environments, where full information about resource richness was provided at the outset—supporting our interpretation that the effect is driven by uncertainty reduction rather than motor or attentional factors. We also found that the order of blocks experienced by participants showed two significant effects. First, participants who performed the long battery block in the first part of the experiment showed a better performance in the second half of the experiment, when they performed the short battery block, compared to participants who performed the short battery block first. This finding suggests that the longer training time enjoyed by participants who performed the long battery block first translates into an advantage in the second block. Second, participants who started the experiment with the short battery block visited significantly more areas in the second (long battery) block. A possible explanation is that during the short battery block, participants learned it was more convenient to leave areas quickly to access the more rewarding boxes and this behavior persisted in the second experimental block. Taken together, these two effects indicate that training time and strategies learned during the fist block influences subsequent performance, highlighting the importance of studying learning dynamics and not only steady-state behavior during foraging experiments. Future studies could assess what sequence of blocks (or training curriculum) could be more advantageous to improve foraging strategies and whether, with sufficient time, human participants could match the performance of optimal agents.

Concerning the third question, we found that participants' performance -- in terms of total coins collected -- differed significantly from an optimal agent in the trials with long battery, but not in those with short battery. The fact that participants in short battery trials approximated better optimal solutions could be due to a combination of (at least) two factors: first, as discussed above, participants are faster in short battery trials; second, both participants and the optimal agent collect fewer coins in short battery trials and the difference might not rise statistical significance. Participants differed significantly from the optimal agent also for the number of boxes collected in the first area (which indexes the flexible adaptation of foraging strategies), during both long battery and short battery trials. While the difference between participants and the optimal agent was not visible when looking at the results of the short battery trials in the mixed environment (Figure \ref{fig:Optimality}), further splitting the mixed environment data into trials in which the first visited area was rich or poor shows that in the former case participants underharvest, whereas in the latter case they overharvest (Figure \ref{fig:OptimalityFirstArea}). The differences between the participants and the optimal agent are not surprising, especially when considering that the optimal agent has perfect knowledge of the environment, whereas participants make decisions under various sources of uncertainty, such as where and when they will be able to find the next box (at at least in the first part of the experiment, their expected content). These differences might translate into more or less deterministic choices. For example, the results of the mixed environment (Figure \ref{fig:OptimalityFirstArea}) showed that the optimal agent privileged extreme solutions, to either collect all the five available boxes (if the area is rich and the battery long) or to collect just one box and leave (if the area is poor and the battery short). Rather, participants selected similar but less extreme solutions, showing some "regression to the mean", which could be explained by participants' uncertainty and/or additional stochasticity in their decision processes. Finally, we found that over the course of the experiment, participants' performance approximated but did not reach the optimal agent (Figure~\ref{fig:optimalityTrials}), suggesting that even with more training, participants' strategies retain some aspects that are not well captured by optimal agents with perfect knowledge. Understanding what causes the differences between participants' and optimal agents' strategies (e.g., heuristics against "extreme" solutions or some rational trade-off) remains to be investigated in future studies.


One limitation of the current study is that we examined foraging decisions using a relatively simple model, which assumed uncertainty only about the locations of rich and poor areas. This approach was chosen because we aimed to investigate how closely people approximate optimal foraging strategies during the experiment, as they progressively learned the task structure—and thus, ideally, remained uncertain only about the locations of rich and poor areas, similar to the optimal model. Future studies could formalize alternative strategies that individuals might adopt when deciding how to forage and when to leave a patch. Previous research has identified possible foraging rules (e.g., giving-up time and area-restricted search in \citep{wilke2009fishing}) that could be adapted to our task and used for model comparison, providing further insight into how people forage and whether their strategies change between early and later phases of foraging. Additionally, incorporating learning models could offer a more detailed examination of explore-exploit dynamics during foraging \citep{friston2016active,sutton2018reinforcement}. Moreover, despite the present study focuses on behavior, our paradigm was informed by neuroscientific models of foraging that implicate brain areas such as the dorsal anterior cingulate cortex and the hippocampus in regulating stay-or-leave decisions, and dopaminergic systems in effort allocation under time constraints \citep{Wittmann2016,LeHeron2019,Ianni2023,fine2025complementary}. Future work could adapt our task to investigate the neural processes supporting flexible foraging with varying resource distribution and limited time. Finally, future studies might adapt our setup to clinical or subclinical populations to investigate under which conditions they develop adaptive or maladaptive foraging strategies \citep{addicott2017primer,barack2024attention,pellicano2011children}.


In sum, our results highlight the great flexibility of human foraging in response to varying resource distributions, limited foraging time, and learning opportunities. Furthermore, they showcase the importance of studying foraging in rich environments that engage various cognitive processes, such as orientation and spatial navigation, while also affording learning and strategy adaptation over time. These ecologically valid contexts enable new questions about how foraging strategies adapt to the physical properties of the environment, revealing subtle aspects often overlooked in accounts like optimal foraging theory. For example, our experiment suggests that people transition from an initial exploration (or information gathering) phase—where they reduce uncertainty about key aspects of the task, such as resource locations and travel times—to a subsequent exploitation (or reward harvesting) phase, in which they leverage their knowledge to deploy more effective foraging strategies. These learning and adaptation effects in foraging present a valuable avenue for future research.



\section*{Declarations}

\subsection*{Funding}
This research received funding from the European Union’s Horizon 2020 Framework Programme for Research and Innovation under the Specific Grant Agreements No. 952215 (TAILOR) to G.P.; the European Research Council under the Grant Agreement No. 820213 (ThinkAhead) to G.P.; the Italian National Recovery and Resilience Plan (NRRP), M4C2, funded by the European Union – NextGenerationEU (Project IR0000011, CUP B51E22000150006, “EBRAINS-Italy”; Project PE0000013, CUP B53C22003630006, "FAIR"; Project PE0000006, CUP J33C22002970002 “MNESYS”) to G.P., PRIN PNRR P20224FESY to G.P., and the National Research Council, project iForagers. The funders had no role in study design, data collection and analysis, decision to publish, or preparation of the manuscript.

\subsection*{Conflicts of interest/Competing interests}
The authors have declared that no competing interests exist.

\subsection*{Ethics approval}
All the procedures were approved by the CNR Ethics committee.

\subsection*{Consent to participate}
Informed consent was obtained from all participants prior to their inclusion in the study. Participants were provided with detailed information about the study's purpose, procedures, and potential risks, and they voluntarily agreed to participate.

\subsection*{Consent for publication}
Informed consent for publication was obtained from all participants. They were provided with detailed information about the study and its intended dissemination, and they voluntarily agreed to the publication of the findings.

\subsection*{Availability of data and materials}
The data that support the findings of this study are available from the corresponding author upon reasonable request.

\subsection*{Code Availability}
Scripts used for the main model, statistical analysis, and result visualization are provided at the following link:
\url{https://github.com/DavideNuzzi/Human-foraging-strategies-flexibly-adapt-to-resource-distribution-and-time-constraints}

\subsection*{Authors' contributions}
All authors contributed to the research and manuscript preparation. Valeria Simonelli was responsible for Conceptualization, Methodology, Data Collection, Data Curation, Formal Analysis, Validation, Visualization, and Writing – Original Draft and Review. Davide Nuzzi contributed to Methodology, Data Curation, Formal Analysis, Validation, Software, Visualization, and Writing – Original Draft and Review. Gian Luca Lancia was involved in Conceptualization, Methodology, and Writing – Review. Giovanni Pezzulo contributed to Conceptualization, Methodology, Supervision, Writing – Original Draft and Review, and Funding Acquisition.

\bibliographystyle{apalike}
\bibliography{references}

\appendix

\setcounter{figure}{0}
\makeatletter 
\renewcommand{\thefigure}{S\@arabic\c@figure}
\makeatother

\setcounter{table}{0}
\makeatletter 
\renewcommand{\thetable}{S\@arabic\c@table}
\makeatother

\newpage

\section{Supplementary analyses}

\subsection{Supplementary tables of the analysis reported in Section \ref{sec:results_all_trials}}

Here, we report the estimated parameters and statistics of linear mixed-effects modeling on six dependent variables: total coins collected, boxes collected in the first area, total boxes collected, average coins per box, number of visited areas, and average box collection time. This analysis considers the effects of our manipulations, namely the battery length, either long or short, and environmental richness, either rich or mixed.

\begin{table}[!htbp]
    \centering
    \small
    \begin{tabular}{lcccccc}
        \hline
        & Coeff. & SE & t & 2.5\_ci & 97.5\_ci & P-val \\
        \hline
        Intercept & 34.703 & 1.205 & 28.791 & 32.340 & 37.065 & $< 0.001$\\
        LongBattery & 28.356 & 0.673 & 42.161 & 27.038 & 29.674 & \bm{$< 0.001$}\\
        RichEnvironment & 7.241 & 0.673 & 10.767 & 5.923 & 8.559 & \bm{$< 0.001$}\\
        LongBattery $\times$ RichEnvironment & 5.600 & 0.951 & 5.888 & 3.736 & 7.464 & \bm{$< 0.001$}\\
        \hline
    \end{tabular}
    \caption{Estimated parameters and statistics of linear mixed-effects modeling on the total number of coins gained per trial.}\label{tab:coinsgainedpertrial_2x2}
\end{table}

\begin{table}[!htbp]
    \centering
    \small
    \begin{tabular}{lcccccc}
        \hline
        & Coeff. & SE & t & 2.5\_ci & 97.5\_ci & P-val \\
        \hline
        Intercept & 2.238 & 0.115 & 19.380 & 2.012 & 2.465 & $< 0.001$\\
        LongBattery & 0.600 & 0.081 & 7.431 & 0.442 & 0.758 & \bm{$< 0.001$}\\
        RichEnvironment & 0.318 & 0.081 & 3.934 & 0.159 & 0.476 & \bm{$< 0.001$}\\
        LongBattery $\times$ RichEnvironment & 0.194 & 0.114 & 1.700 & -0.030 & 0.418 & 0.089\\
        \hline
    \end{tabular}
    \caption{Estimated parameters and statistics of linear mixed-effects modeling on the number of boxes collected in the first area.}
    \label{tab:boxopenedinthefirstarea_2x2}
\end{table}

\begin{table}[!htbp]
    \centering
    \small
    \begin{tabular}{lcccccc}
        \hline
        & Coeff. & SE & t & 2.5\_ci & 97.5\_ci & P-val \\
        \hline
        Intercept & 5.582 & 0.206 & 27.057 & 5.178 & 5.987 & $< 0.001$\\
        LongBattery & 5.432 & 0.113 & 48.045 & 5.211 & 5.654 & \bm{$< 0.001$}\\
        RichEnvironment & 0.244 & 0.113 & 2.159 & 0.023 & 0.466 & \bm{$0.031$}\\
        LongBattery $\times$ RichEnvironment & -0.018 & 0.160 & -0.110 & -0.331 & 0.296 & 0.912\\
        \hline
    \end{tabular}
    \caption{Estimated parameters and statistics of linear mixed-effects modeling on the number of boxes opened per trial.}
    \label{tab:boxopenedpertrial_2x2}
\end{table}

\begin{table}[!htbp]
    \centering
    \small
    \begin{tabular}{lcccccc}
        \hline
        & Coeff. & SE & t & 2.5\_ci & 97.5\_ci & P-val \\
        \hline
        Intercept & 6.231 & 0.044 & 143.214 & 6.146 & 6.316 & $< 0.001$\\
        LongBattery & -0.454 & 0.032 & -14.167 & -0.391 & 0.851 & \bm{$< 0.001$}\\
        RichEnvironment & 0.999 & 0.032 & 31.182 & 0.936 & 1.061 & \bm{$< 0.001$}\\
        LongBattery $\times$ RichEnvironment & 0.002 & 0.045 & 0.042 & -0.087 & 0.091 & 0.966\\
        \hline
    \end{tabular}
    \caption{Estimated parameters and statistics of linear mixed-effects modeling on the average coins collected per box.}\label{tab:coinsgainedpertrialmean_2x2}
\end{table}

\begin{table}[!htbp]
    \centering
    \small
    \begin{tabular}{lcccccc}
        \hline
        & Coeff. & SE & t & 2.5\_ci & 97.5\_ci & P-val \\
        \hline
        Intercept & 2.965 & 0.081 & 36.730 & 2.807 & 3.123 & $< 0.001$\\
        LongBattery & 0.765 & 0.044 & 17.342 & 0.678 & 0.851 & \bm{$< 0.001$}\\
        RichEnvironment & -0.050 & 0.044 & -1.134 & -0.136 & 0.036 & 0.257\\
        LongBattery $\times$ RichEnvironment & -0.015 & 0.062 & -0.236 & -0.137 & 0.108 & 0.814\\
        \hline
    \end{tabular}
     \caption{Estimated parameters and statistics of linear mixed-effects modeling on the number of visited areas.}
    \label{tab:visited_areas_2x2}
\end{table}

\begin{table}[!htbp]
    \centering
    \small
    \begin{tabular}{lcccccc}
        \hline
        & Coeff. & SE & t & 2.5\_ci & 97.5\_ci & P-val \\
        \hline
        Intercept & 3.324 & 0.109 & 30.413 & 3.110 & 3.539 & $< 0.001$\\
        LongBattery & 0.263 & 0.073 & 3.597 & 0.120 & 0.406 & \bm{$< 0.001$}\\
        RichEnvironment & -0.022 & 0.073 & -0.300 & -0.165 & 0.121 & 0.764\\
        LongBattery $\times$ RichEnvironment & -0.014 & 0.103 & -0.136 & -0.216 & 0.188 & 0.892\\
        \hline
    \end{tabular}
    \caption{Estimated parameters and statistics of linear mixed-effects modeling on the average box collection time.}\label{tab:averageboxcollectiontime_2x2}
\end{table}

\begin{table}[htb]
\centering
\begin{tabular}{lccccc}
\toprule
Environment & Battery & Contrast & Estimate & Std Error & p-value\\
\hline
Mixed & . & Short - Long & -28.356 & 0.673 & \bm{$< 0.001$}\\
Rich & . & Short - Long & -33.956 & 0.673 & \bm{$< 0.001$}\\
. & Short & Mixed - Rich & -7.241 & 0.673 & \bm{$< 0.001$}\\
. & Long & Mixed - Rich & -12.841 & 0.673 & \bm{$< 0.001$}\\
\bottomrule
\end{tabular}
\caption{Pairwise post-hoc comparisons for the variable: total coins}
\label{tab:Comparisons_CoinsGainedPerTrial}
\end{table}

\begin{table}[htb]
\centering
\begin{tabular}{lccccc}
\toprule
Environment & Battery & Contrast & Estimate & Std Error & p-value\\
\hline
Mixed & . & Short - Long & -0.600 & 0.081 & \bm{$< 0.001$}\\
Rich & . & Short - Long & -0.794 & 0.081 & \bm{$< 0.001$}\\
. & Short & Mixed - Rich & -0.318 & 0.081 & \bm{$< 0.001$}\\
. & Long & Mixed - Rich & -0.512 & 0.081 & \bm{$< 0.001$}\\
\bottomrule
\end{tabular}
\caption{Pairwise post-hoc comparisons for the variable: boxes collected in first area}
\label{tab:Comparisons_BoxOpenedFirstArea}
\end{table}

\begin{table}[htb]
\centering
\begin{tabular}{lccccc}
\toprule
Environment & Battery & Contrast & Estimate & Std Error & p-value\\
\hline
Mixed & . & Short - Long & -5.432 & 0.113 & \bm{$< 0.001$}\\
Rich & . & Short - Long & -5.415 & 0.113 & \bm{$< 0.001$}\\
. & Short & Mixed - Rich & -0.244 & 0.113 & 0.118\\
. & Long & Mixed - Rich & -0.226 & 0.113 & 0.170\\
\bottomrule
\end{tabular}
\caption{Pairwise post-hoc comparisons for the variable: boxes collected}
\label{tab:Comparisons_BoxOpenedPerTrial}
\end{table}

\begin{table}[htb]
\centering
\begin{tabular}{lccccc}
\toprule
Environment & Battery & Contrast & Estimate & Std Error & p-value\\
\hline
Mixed & . & Short - Long & 0.454 & 0.032 & \bm{$< 0.001$}\\
Rich & . & Short - Long & 0.452 & 0.032 & \bm{$< 0.001$}\\
. & Short & Mixed - Rich & -0.999 & 0.032 & \bm{$< 0.001$}\\
. & Long & Mixed - Rich & -1.001 & 0.032 & \bm{$< 0.001$}\\
\bottomrule
\end{tabular}
\caption{Pairwise post-hoc comparisons for the variable: average coins per box}
\label{tab:Comparisons_CoinGainedPerBoxPerTrialMean}
\end{table}

\begin{table}[htb]
\centering
\begin{tabular}{lccccc}
\toprule
Environment & Battery & Contrast & Estimate & Std Error & p-value\\
\hline
Mixed & . & Short - Long & -0.765 & 0.044 & \bm{$< 0.001$}\\
Rich & . & Short - Long & -0.750 & 0.044 & \bm{$< 0.001$}\\
. & Short & Mixed - Rich & 0.050 & 0.044 & 0.695\\
. & Long & Mixed - Rich & 0.065 & 0.044 & 0.459\\
\bottomrule
\end{tabular}
\caption{Pairwise post-hoc comparisons for the variable: number of visited areas}
\label{tab:Comparisons_VisitedAreasEnteredOnce}
\end{table}

\begin{table}[htb]
\centering
\begin{tabular}{lccccc}
\toprule
Environment & Battery & Contrast & Estimate & Std Error & p-value\\
\hline
Mixed & . & Short - Long & -0.263 & 0.073 & 0.001\\
Rich & . & Short - Long & -0.249 & 0.073 & 0.003\\
. & Short & Mixed - Rich & 0.022 & 0.073 & 0.997\\
. & Long & Mixed - Rich & 0.036 & 0.073 & 0.980\\
\bottomrule
\end{tabular}
\caption{Pairwise post-hoc comparisons for the variable: time between boxes}
\label{tab:Comparisons_TempiTraCasse}
\end{table}

\newpage
\FloatBarrier

\subsection{Supplementary tables of the analysis reported in Section \ref{sec:mixed_env_analysis}}

In this section, we present the estimated parameters and statistics of linear mixed-effects modeling on six dependent variables: total coins collected, boxes collected in the first area, total boxes collected, average coins per box, number of visited areas, and average box collection time. The analysis considers the effect of our manipulations taking into account the richness of the first area visited by participants, in mixed environment trials.

\begin{table}[!htbp]
    \centering
    \small
    \begin{tabular}{lcccccc}
        \hline
        & Coeff. & SE & t & 2.5\_ci & 97.5\_ci & P-val \\
        \hline
        Intercept & 333.732 & 1.122 & 30.063 & 31.533 & 35.931 & $< 0.001$\\
        LongBattery & 29.008 & 0.836 & 34.716 & 27.370 & 30.646 & \bm{$< 0.001$}\\
        FirstAreaRich & 1.931 & 0.822 & 2.349 & 0.320 & 3.542 & \bm{$0.019$}\\
        LongBattery $\times$ FirstAreaRich & -1.345 & 1.159 & -1.161 & -3.616 & 0.926 & 0.246\\
        \hline
    \end{tabular}
    \caption{Estimated parameters and statistics of linear mixed-effects modeling on the total number of coins gained per trial.} \label{coinsgainedpertrial_2x2x2}
\end{table}

\begin{table}[!htbp]
    \centering
    \small
    \begin{tabular}{lcccccc}
        \hline
        & Coeff. & SE & t & 2.5\_ci & 97.5\_ci & P-val \\
        \hline
        Intercept & 1.632 & 0.115 & 14.241 & 1.408 & 1.857 & $< 0.001$\\
        LongBattery & 0.290 & 0.100 & 2.893 & 0.093 & 0.486 & \bm{$0.004$}\\
        FirstAreaRich & 1.205 & 0.098 & 12.239 & 1.012 & 1.398 & \bm{$< 0.001$}\\
        LongBattery $\times$ FirstAreaRich & 0.479 & 0.139 & 3.451 & 0.207 & 0.751 & \bm{$0.001$}\\
        \hline
    \end{tabular}
    \caption{Estimated parameters and statistics of linear mixed-effects modeling on the number of boxes collected in the first area.}\label{tab:boxopenedfirstarea_2x2x2}
\end{table}

\begin{table}[!htbp]
    \centering
    \small
    \begin{tabular}{lcccccc}
        \hline
        & Coeff. & SE & t & 2.5\_ci & 97.5\_ci & P-val \\
        \hline
        Intercept & 5.606 & 0.217 & 25.779 & 5.180 & 6.032 & $< 0.001$\\
        LongBattery & 5.404 & 0.159 & 33.997 & 5.092 & 5.715 & \bm{$< 0.001$}\\
        FirstAreaRich & -0.047 & 0.156 & -0.301 & -0.354 & 0.259 & 0.763\\
        LongBattery $\times$ FirstAreaRich & 0.056 & 0.220 & 0.255 & -0.376 & 0.488 & 0.799\\
        \hline
    \end{tabular}
    \caption{Estimated parameters and statistics of linear mixed-effects modeling on the number of boxes opened per trial.} \label{tab:boxopenedpertrial_2x2x2}
\end{table}

\begin{table}[!htbp]
    \centering
    \small
    \begin{tabular}{lcccccc}
        \hline
        & Coeff. & SE & t & 2.5\_ci & 97.5\_ci & P-val \\
        \hline
        Intercept & 6.002 & 0.053 & 112.410 & 5.898 & 6.107 & $< 0.001$\\
        LongBattery & -0.263 & 0.053 & -4.953 & -0.367 & -0.159 & \bm{$< 0.001$}\\
        FirstAreaRich & 0.455 & 0.052 & 8.724 & 0.353 & 0.557 & \bm{$< 0.001$}\\
        LongBattery $\times$ FirstAreaRich & -0.385 & -0.241 & -5.242 & -0.530 & -0.241 & \bm{$< 0.000$}\\
        \hline
    \end{tabular}
    \caption{Estimated parameters and statistics of linear mixed-effects modeling on the average coins collected per box.}\label{tab:coinsgainedpertrialmean_2x2x2}
\end{table}

\begin{table}[!htbp]
    \centering
    \small
    \begin{tabular}{lcccccc}
        \hline
        & Coeff. & SE & t & 2.5\_ci & 97.5\_ci & P-val \\
        \hline
        Intercept & 3.129 & 0.082 & 38.353 & 2.969 & 3.289 & $< 0.001$\\
        LongBattery & 0.718 & 0.064 & 11.199 & 0.592 & 0.844 & \bm{$< 0.001$}\\
        FirstAreaRich & -0.326 & 0.063 & -5.171 & -0.450 & -0.202 & \bm{$< 0.001$}\\
        LongBattery $\times$ FirstAreaRich & 0.111 & 0.089 & 1.245 & -0.064 & 0.285 & 0.214\\
        \hline
    \end{tabular}
    \caption{Estimated parameters and statistics of linear mixed-effects modeling on the number of visited areas.}
    \label{tab:visited_areas_2x2x2}
\end{table}

\begin{table}[!htbp]
    \centering
    \small
    \begin{tabular}{lcccccc}
        \hline
        & Coeff. & SE & t & 2.5\_ci & 97.5\_ci & P-val \\
        \hline
        Intercept & 3.223 & 0.119 & 27.023 & 2.989 & 3.456 & $< 0.001$\\
        LongBattery & 0.291 & 0.109 & 2.679 & 0.078 & 0.505 & \bm{$0.008$}\\
        FirstAreaRich & 0.202 & 0.107 & 1.889 & -0.008 & 0.412 & 0.059\\
        LongBattery $\times$ FirstAreaRich & -0.068 & 0.151 & -0.452 & -0.364 & 0.228 & 0.652\\
        \hline
    \end{tabular}
    \caption{Estimated parameters and statistics of linear mixed-effects modeling on the average box collection time.} \label{tab:averageboxcollectiontime_2x2x2}
\end{table}

\begin{table}[htb]
\centering
\begin{tabular}{lccccc}
\toprule
First Area & Battery & Contrast & Estimate & Std Error & p-value\\
\hline
Poor & . & Short - Long & -29.008 & 0.836 & \bm{$< 0.001$}\\
Rich & . & Short - Long & -27.663 & 0.797 & \bm{$< 0.001$}\\
. & Short & Poor - Rich & -1.931 & 0.822 & 0.074\\
. & Long & Poor - Rich & -0.586 & 0.825 & 0.926\\
\bottomrule
\end{tabular}
\caption{Pairwise post-hoc comparisons for the variable: total coins}
\label{tab:ComparisonsFirstRich_CoinsGainedPerTrial}
\end{table}

\begin{table}[htb]
\centering
\begin{tabular}{lccccc}
\toprule
First Area & Battery & Contrast & Estimate & Std Error & p-value\\
\hline
Poor & . & Short - Long & -0.290 & 0.100 & 0.016\\
Rich & . & Short - Long & -0.769 & 0.095 & \bm{$< 0.001$}\\
. & Short & Poor - Rich & -1.205 & 0.098 & \bm{$< 0.001$}\\
. & Long & Poor - Rich & -1.684 & 0.099 & \bm{$< 0.001$}\\
\bottomrule
\end{tabular}
\caption{Pairwise post-hoc comparisons for the variable: boxes collected in first area}
\label{tab:ComparisonsFirstRich_BoxOpenedFirstArea}
\end{table}

\begin{table}[htb]
\centering
\begin{tabular}{lccccc}
\toprule
First Area & Battery & Contrast & Estimate & Std Error & p-value\\
\hline
Poor & . & Short - Long & -5.404 & 0.159 & \bm{$< 0.001$}\\
Rich & . & Short - Long & -5.460 & 0.152 & \bm{$< 0.001$}\\
. & Short & Poor - Rich & 0.047 & 0.156 & 0.997\\
. & Long & Poor - Rich & -0.009 & 0.157 & 1.000\\
\bottomrule
\end{tabular}
\caption{Pairwise post-hoc comparisons for the variable: boxes collected}
\label{tab:ComparisonsFirstRich_BoxOpenedPerTrial}
\end{table}

\begin{table}[htb]
\centering
\begin{tabular}{lccccc}
\toprule
First Area & Battery & Contrast & Estimate & Std Error & p-value\\
\hline
Poor & . & Short - Long & 0.263 & 0.053 & \bm{$< 0.001$}\\
Rich & . & Short - Long & 0.648 & 0.051 & \bm{$< 0.001$}\\
. & Short & Poor - Rich & -0.455 & 0.052 & \bm{$< 0.001$}\\
. & Long & Poor - Rich & -0.069 & 0.052 & 0.560\\
\bottomrule
\end{tabular}
\caption{Pairwise post-hoc comparisons for the variable: average coins per box}
\label{tab:ComparisonsFirstRich_CoinGainedPerBoxPerTrialMean}
\end{table}

\begin{table}[htb]
\centering
\begin{tabular}{lccccc}
\toprule
First Area & Battery & Contrast & Estimate & Std Error & p-value\\
\hline
Poor & . & Short - Long & -0.718 & 0.064 & \bm{$< 0.001$}\\
Rich & . & Short - Long & -0.829 & 0.061 & \bm{$< 0.001$}\\
. & Short & Poor - Rich & 0.326 & 0.063 & \bm{$< 0.001$}\\
. & Long & Poor - Rich & 0.215 & 0.063 & 0.003\\
\bottomrule
\end{tabular}
\caption{Pairwise post-hoc comparisons for the variable: number of visited areas}
\label{tab:ComparisonsFirstRich_VisitedAreasEnteredOnce}
\end{table}

\begin{table}[htb]
\centering
\begin{tabular}{lccccc}
\toprule
First Area & Battery & Contrast & Estimate & Std Error & p-value\\
\hline
Poor & . & Short - Long & -0.291 & 0.109 & 0.030\\
Rich & . & Short - Long & -0.223 & 0.104 & 0.121\\
. & Short & Poor - Rich & -0.202 & 0.107 & 0.217\\
. & Long & Poor - Rich & -0.134 & 0.107 & 0.615\\
\bottomrule
\end{tabular}
\caption{Pairwise post-hoc comparisons for the variable: time between boxes}
\label{tab:ComparisonsFirstRich_TempiTraCasse}
\end{table}

\FloatBarrier
\newpage

\subsection{Supplementary analysis, restricted to trials in which the first visited area was rich, in both rich and mixed environments}

\begin{figure}[!htbp]
    \centering
    \includegraphics[width=0.5\linewidth]{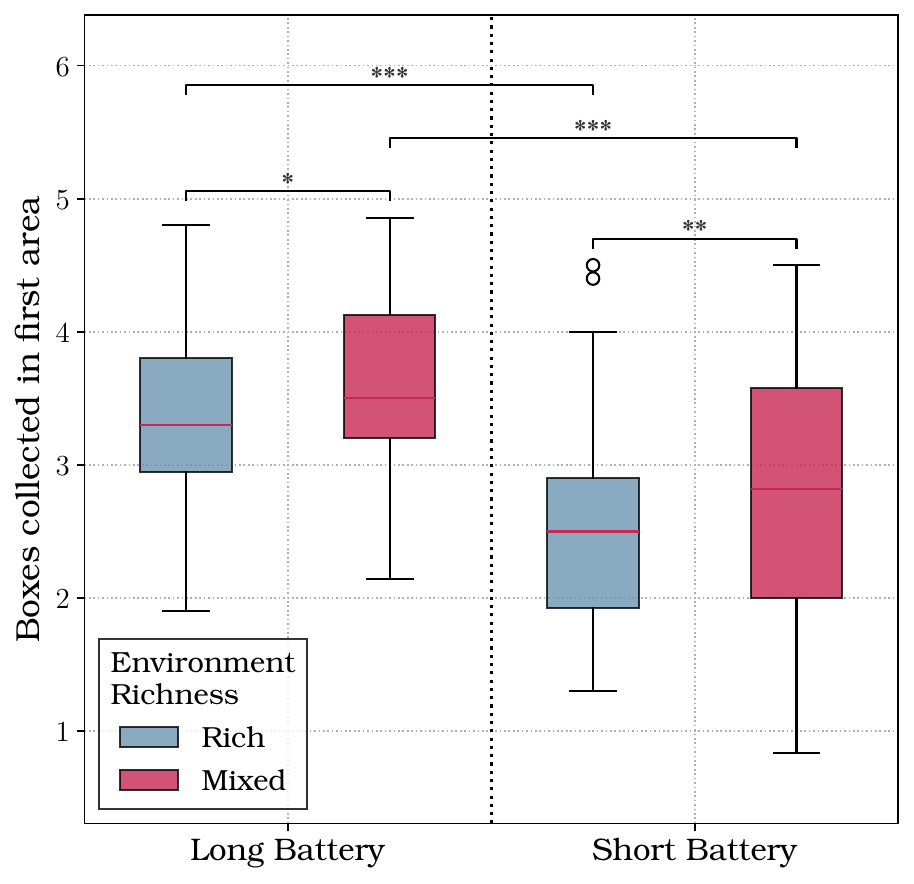}
    \caption{Boxes collected in the first area, only when it is rich, for both rich and mixed environments. See the main text for explanation.}
    \label{fig:boxplotfirstarearich_boxesfirst}
\end{figure}

In this section, we present the estimated parameters and statistics of linear mixed-effects modeling on boxes collected in the first area. The analysis considers the effect of our manipulations, namely the battery length, either long or short, and environmental richness, either rich or mixed, but is restricted to cases in which the first visited area was rich (which is always true in the rich environment, but only approximately half of the times in the mixed environment). See also Figure \ref{fig:boxplotfirstarearich_boxesfirst}.

\begin{table}[!htbp]
    \centering
    \small
    \begin{tabular}{lcccccc}
        \hline
        & Coeff. & SE & t & 2.5\_ci & 97.5\_ci & P-val \\
        \hline
        Intercept & 2.842 & 0.131 & 21.621 & 2.585 & 3.100 & $< 0.001$\\
        LongBattery & 0.766 & 0.097 & 7.884 & 0.576 & 0.957 & \bm{$< 0.001$}\\
        RichEnvironment & -0.286 & 0.086 & -3.331 & -0.455 & -0.118 & \bm{$0.001$}\\
        LongBattery $\times$ RichEnvironment & 0.038 & 0.120 & 0.315 & -0.197 & 0.273 & 0.753\\
        \hline
    \end{tabular}
    \caption{Estimated parameters and statistics of linear mixed-effects modeling on the number of boxes collected in the first area.}\label{tab:boxopenedfirstarearich}
\end{table}

\newpage

\subsection{Supplementary Figures}

\begin{figure}[!htbp]
    \centering
    \includegraphics[width=0.9\linewidth]{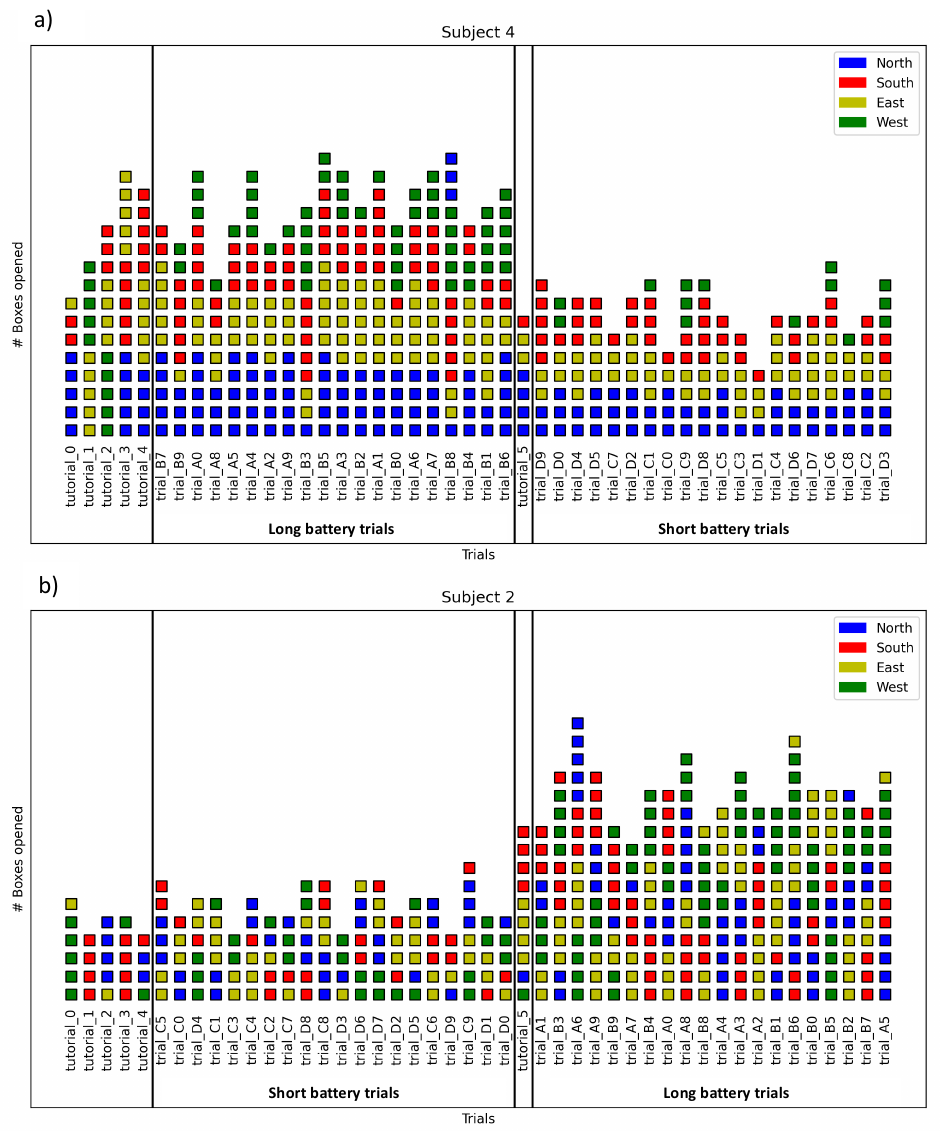}
    \caption{Box-opening patterns throughout the experiment for two representative participants employing different navigational strategies. Each square represents a box opened, with colors indicating the area to which the box belonged. (a) Subject 4 follows a stable, clockwise navigational pattern, while (b) Subject 2 exhibits an unstable navigational pattern, highlighting differences in area preferences and exploration strategies over time. See the main text for explanation.}
    \label{fig:navigationalpatterns}
\end{figure}

\begin{figure}[!htbp]
    \centering
    \includegraphics[width=0.8\linewidth]{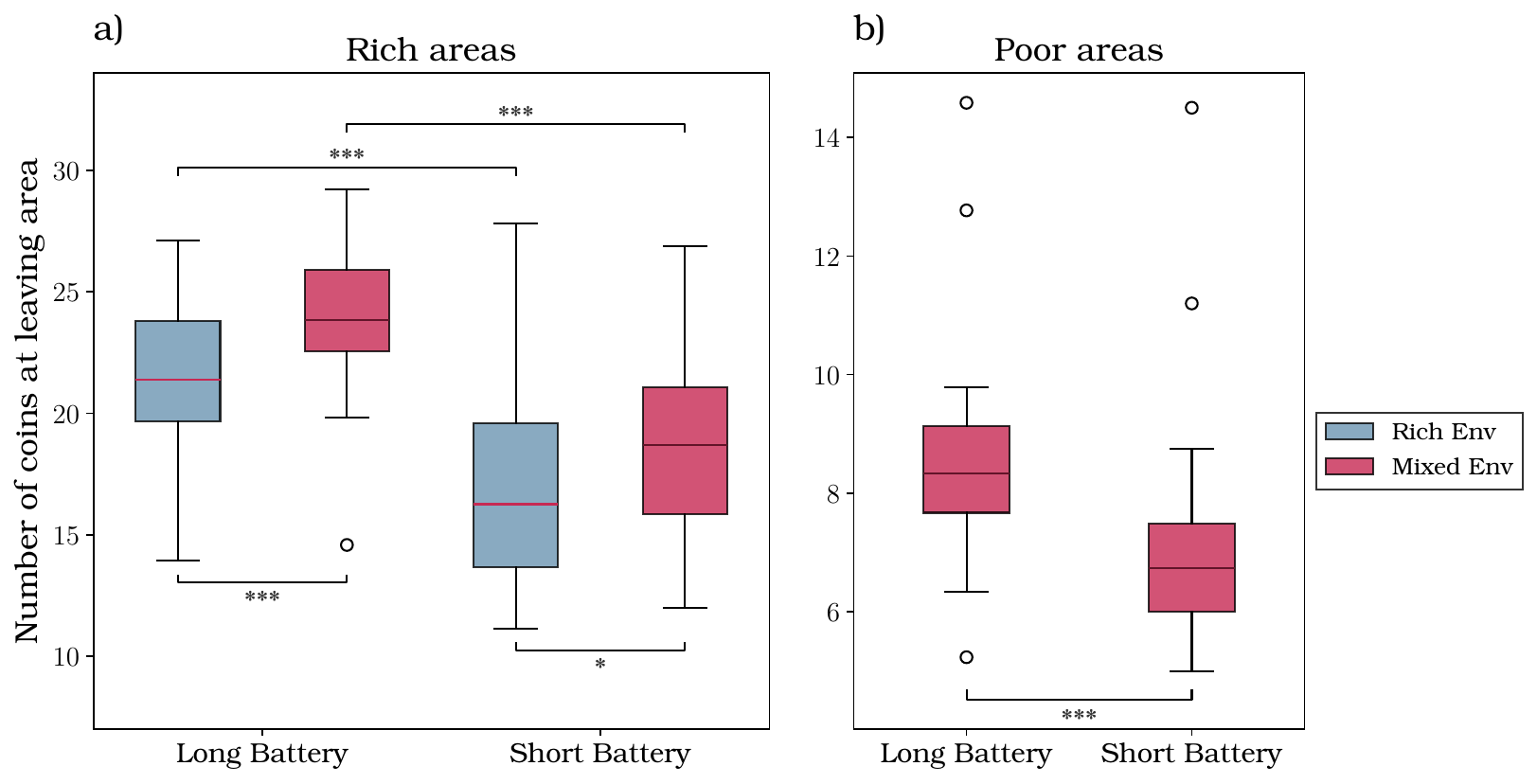}
    \caption{Number of coins collected by participants before leaving (a) rich or (b) poor areas. The number of collected coins changes significantly across conditions ($p = 0.041$ for the environment richness comparison in the short battery condition, $p < 0.001$ in all the other cases, Mann-Whitney U-test, FDR correction), suggesting that participants do not adopt a threshold-driven strategy to decide when to leave areas. See the main text for explanation.}
    \label{fig:CoinsForArea}
\end{figure}

\begin{figure}[!htbp]
    \centering
    \includegraphics[width=0.5\linewidth]{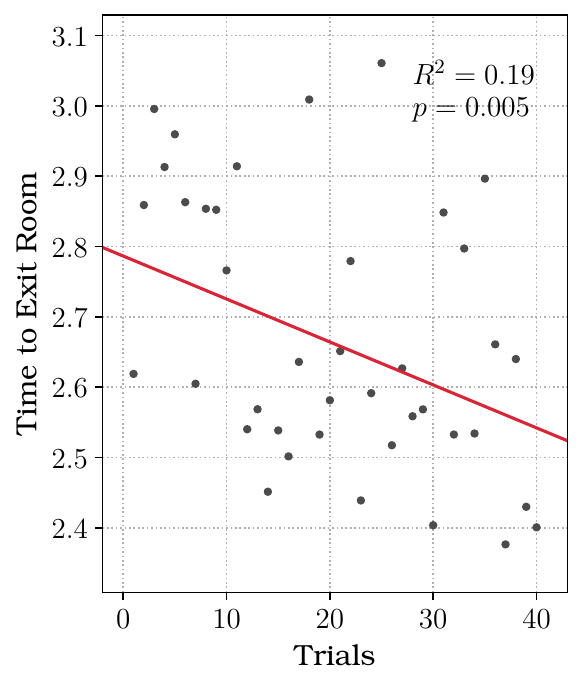}
    \caption{Participants' \emph{exit time}: average time to leave a room after collecting the last treasure box within it, across different experimental conditions. The red line represents the linear regression fit, highlighting a significant decrease in time to exit as participants progressed through the trials (R² = 0.19, p = 0.005). See the main text for explanation.}
    \label{fig:learningExit}
\end{figure}

\begin{figure}[!htbp]
    \centering
    \includegraphics[width=0.5\linewidth]{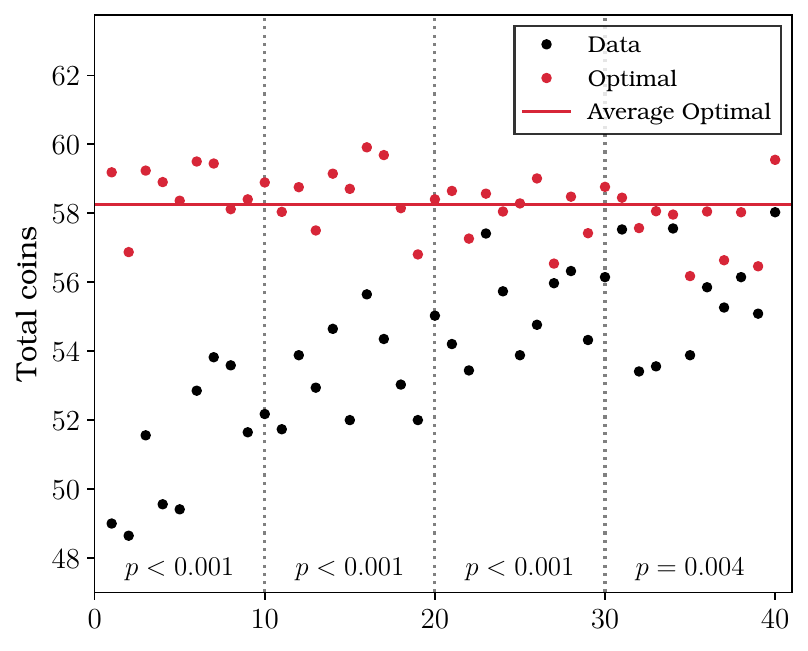}
    \caption{Participants' performance increases over time, approximating but not reaching the optimal agent. The plot compares the total coins earned by the optimal agent (red line is the average) and participants (black dots). Note that participants' data are the same as Figure~\ref{fig:learning}.a. The results of four t-tests (shown at the bottom of the figure) show that the performance of the optimal agent is significantly better, across all the four periods considered. See the main text for explanation.}
    \label{fig:optimalityTrials}
\end{figure}

\end{document}